\documentclass[11pt]{article}

\usepackage{graphicx}
\usepackage{slashed}

\usepackage{epsfig} 
\usepackage{graphics}
\newcommand{\sect}[1]{ \section{#1} \setcounter{equation}{0} } 

\newcommand{\half}{\mbox{\small{$\frac{1}{2}$}}} 
 
\newcommand{\sixth}{\mbox{\small{$\frac{1}{6}$}}} 
\newcommand{\MSbar}{\overline{\mbox{MS}}} 
\newcommand{\MSbars}{\overline{\mbox{\footnotesize{MS}}}} 
\newcommand{\RI}{\mbox{RI${}^\prime$}}
\newcommand{\RIs}{\mbox{\footnotesize{RI${}^\prime$}}}
\newcommand{\MOMc}{\mbox{MOMc}}
\newcommand{\MOMg}{\mbox{MOMg}}
\newcommand{\MOMq}{\mbox{MOMq}}
\newcommand{\mMOM}{\mbox{mMOM}}
\newcommand{\MOMcs}{\mbox{\footnotesize{MOMc}}}
\newcommand{\MOMgs}{\mbox{\footnotesize{MOMg}}}
\newcommand{\MOMqs}{\mbox{\footnotesize{MOMq}}}
\newcommand{\mMOMs}{\mbox{\footnotesize{mMOM}}}

\newcommand{\MOMi}{\mbox{MOMi}}

\newcommand{\Nf}{N_{\!f}}
\newcommand{\NF}{N_{\!F}}
\newcommand{\psid}{\psi^{(1)}({\mbox{\small{$\frac{1}{3}$}}})}
\newcommand{\psiddd}{\psi^{(3)}({\mbox{\small{$\frac{1}{3}$}}})}
\newcommand{\psiddddd}{\psi^{(5)}({\mbox{\small{$\frac{1}{3}$}}})}
\newcommand{\pitwo}{\mbox{\small{$\frac{\pi}{2}$}}}

\newcommand{\pisix}{\mbox{\small{$\frac{\pi}{6}$}}}

\begin{document}

\title{Kinematic scheme study of the $O(a^4)$ Bjorken sum rule and R~ratio}

\author{R.H. Mason \& J.A. Gracey, \\ Theoretical Physics Division, \\
Department of Mathematical Sciences, \\ University of Liverpool, \\ P.O. Box
147, \\ Liverpool, \\ L69 3BX, \\ United Kingdom.}

\date{}

\maketitle

\vspace{5cm}
\noindent
{\bf Abstract.} The Bjorken sum rule and R ratio are constructed to $O(a^4)$ in
the Landau gauge in the three momentum subtraction schemes of Celmaster and 
Gonsalves where $a$~$=$~$g^2/(16\pi^2)$. We aim to examine the issue of 
convergence for observables in the various schemes as well as to test ideas on 
whether using the discrepancy in different scheme values is a viable and more 
quantum field theoretic alternative to current ways of estimating the theory 
error on a measureable.

\vspace{-16.0cm}
\hspace{13.4cm}
{\bf LTH 1352}

\newpage

\sect{Introduction.}

Latterly the development of new methods to evaluate massless Feynman graphs has
led to the computation of various important quantities in Quantum 
Chromodynamics (QCD) to very high loop order. Aside from the recent 
determination of the five loop QCD $\beta$-function in the modified minimal 
subtraction ($\MSbar$) scheme, \cite{1,2,3,4,5}, that underlies the behaviour 
of the gauge coupling constant $g$, the extension of the Bjorken sum rule, 
\cite{6,7,8,9}, and the R ratio, \cite{10,11,12,13,14,15,16,17,18,19}, to high 
loop order has led to a greater precision for comparing to experiment and 
improving estimates of $\alpha_s(M_Z)$ for example. Here $M_Z$ is the mass of 
the $Z$ boson and $\alpha_s$ is the strong coupling constant with
$\alpha_s$~$=$~$g^2/(4\pi)$. This is important at present since the improvement
in collider data necessarily requires that quantum field theory precision has 
to develop in parallel. Despite the progress in Feynman integral evaluation in 
general, the error in measurements is invariably dominated by theory 
uncertainty. From a theoretical point of view the perturbative expansions at 
high loop order are always advanced first in the $\MSbar$ scheme. This is 
primarily due to the fact that one only requires the poles with respect to the 
regularizing parameter to determine the renormalization constants. However, the 
scheme has a drawback in that it is not a kinematic one. By this we mean that
provided the subtraction point where the $\MSbar$ renormalization constants are
defined is not problematic, such as inadvertently and incorrectly introducing 
infrared singularities, then it carries no information associated with the 
kinematics of the subtraction point. Conceptually in an experiment one makes a 
measurement of say the interaction strength at a specific momentum 
configuration. The value recorded there can be tied to the value of the 
coupling constant in the underlying quantum field theory. By the same token one
can define, in a parallel sense, the renormalization of the coupling constant 
at a particular subtraction point. Then a kinematic scheme is constructed 
through a vertex function by defining the coupling constant renormalization 
constant in such a way that after renormalization the vertex function is 
precisely the renormalized coupling constant. In other words the finite part of
the vertex function at the subtraction point momentum configuration is fully 
absorbed into the renormalization constant. This differs from the $\MSbar$ 
scheme where the finite part is ignored in defining the coupling constant 
renormalization constant. 

A set of such schemes was introduced in QCD in the work of Celmaster and 
Gonsalves, \cite{20,21}. In those articles three momentum subtraction (MOM) 
schemes were constructed each based on the three $3$-point vertices of the QCD 
Lagrangian. They were denoted by MOMg, MOMc and MOMq and based respectively on 
the triple gluon, ghost-gluon and quark-gluon vertices. More precisely the 
kinematic configuration considered in \cite{20,21} was the fully symmetric 
point where the squared momenta of the three external legs of the respective 
vertices were equal. In addition in \cite{12,13} the R ratio was computed in 
each MOM scheme to ascertain the respective behaviours. While the low loop 
order studied in \cite{12,13} was not high enough to come to concrete 
conclusions there has been a renewed interest in the development and 
application of MOM schemes to the behaviour of observables. This has been 
brought about via the development of the Laporta algorithm \cite{22}. That 
technique opens the way to evaluate two and three loop $3$-point vertex
functions which can be achieved once the underlying master Feynman integrals
are known. Consequently the three QCD MOM $\beta$-functions were calculated
first to three loops in \cite{23} and then more recently to four loops in
\cite{24}. The former was an exact computation in the sense that the masters 
were known analytically. Earlier work on the MOM $\beta$-functions at that 
order had been numerical \cite{25}. There the masters were determined by
applying the {\sc Mincer} algorithm, \cite{26}, to the momentum expansion of 
each master to high numerical precision before resumming. The remarkable 
accuracy of \cite{25} was apparent once the analytic result became available. A
similar approach was adapted to extract the three loop symmetric point masters 
again using {\sc Mincer}, \cite{24}. Exploiting the {\tt PSLQ} algorithm 
\cite{27} allowed for the translation of the highly precise numerical values of
the masters, derived with {\sc Mincer} and other methods, to analytic functions
whose arguments were the sixth roots of unity, \cite{24}. It is known that such
constants have a connection to cyclotomic polynomials, \cite{28}. One 
consequence of \cite{23} was that the R ratio was determined in the MOM schemes
to $O(a^3)$ in \cite{29}, where $a$~$=$~$g^2/(16\pi^2)$, thereby extending 
\cite{12,13}. This was an instance where the behaviour of an observable could 
be compared in kinematic schemes to those in non-kinematic ones such as 
$\MSbar$. Indeed another non-kinematic one was also considered which was the 
minimal MOM (mMOM) scheme. It was introduced in \cite{30} and is based on the 
ghost-gluon vertex where the momentum of one of the external ghosts is 
nullified. In essence it endeavours to preserve Taylor's observation that the 
ghost-gluon vertex is finite in the Landau gauge, \cite{31}, for other
covariant linear gauges. The $\mMOM$ renormalization group functions have been 
determined to high loop order, \cite{30,32,33}. For the purely theoretical 
situation where quarks are massless as well as overlooking resonances and so 
forth the R ratio behaviour was different in the various schemes. 

One concern was that with the predominant use of the $\MSbar$ scheme 
expressions in error analyses there was potentially another source of theory 
discrepancy lurking in the scheme variations. This is primarily due to the 
truncation of a series which invariably misses information. To address this 
scheme issue various methods have been developed and used to significantly 
improve the truncation uncertainty. There have been several main approaches and
we draw attention to the more popular ones where the associated references are 
not an exhaustive nor definitive list. Rather they are a rough signpost to 
recent literature. For instance a widely used method to estimate higher order 
corrections or theoretical uncertainties is to use the conventional scale 
setting method. In this case the theoretical error at a particular momentum $Q$
is estimated by the maximum and mimimum values of the measureable in the range
$[\half Q, 2 Q]$ at the highest available loop order of a particular scheme. 
One drawback of this is that the value at $\half Q$ may be at a point outside 
the range of perturbative reliability. By contrast the method known as the 
Principle of Maximal Conformality (PMC) has been developed in various 
directions, \cite{34}, and widely applied to several observables. It has a more
field theoretic origin and has been shown to reduce the scale and scheme 
uncertainties significantly. For a recent comprehensive review see, for 
instance, \cite{35}. In the context of this article several current studies are 
worth noting. For example, the PMC was applied to kinematic schemes in 
\cite{36} where the role of the covariant gauge parameter, $\alpha$, was 
included in the analysis. While the $\MSbar$ $\beta$-function does not depend 
on that parameter, \cite{37}, this is not the case in MOM schemes as the 
explicit $\alpha$ dependence for the three MOM $\beta$-functions is available, 
\cite{20,21,23}. Such PMC studies have shown interesting properties. For 
instance, in \cite{38} a PMC study of the $V$-scheme, \cite{39,40}, shows that 
it has advantageous properties compared to the $\MSbar$ scheme. Other 
approaches include the Principle of Minimal Sensitivity, \cite{41,42,43}, 
effective charges, \cite{44,45,46}, the Brodsky-Lepage-Mackenzie method, 
\cite{47} as well as the more recent development of the Principle of Observable
Effective Matching, \cite{48}. A comprehensive review of the scale setting 
problem in QCD can be found, for instance, in \cite{49}. Additionally there are
other approaches such as those that extract information about the higher order 
terms in the perturbative expansion of various quantities. A recent study that 
employs such a technique can be found in \cite{50} for example. Rather than 
attempt to fix the scheme or scale to mitigate residual theoretical 
uncertainties, in this study we attempt to understand and quantify the scheme 
dependence in high loop calculations to improve our understanding of the 
parameterization of theory error in these terms, a technique that is applied in
\cite{51,52}. Moreover while our investigation will be at a more theoretical 
level it is worth noting that the interplay of schemes with error analysis has
already been examined in a more phenomenological context. See, for example, 
\cite{53} for a electroweak sector study as well as, for instance,
\cite{54,55,56} where the extraction of the top quark mass was considered in 
the top mass scheme. A more recent study \cite{57} employed renormalization 
group summed perturbation theory to explore convergence and scale dependence in
the $\mMOM$ scheme for the R ratio and Higgs boson decay.

While such probes of the higher order behaviour have improved our understanding
of experimental results an equally useful way is to determine as far as 
possible the highest order calculable in the perturbative expansion in a 
variety of schemes to see if the scheme dependence shows signs of being washed 
out as the loop order increases. That is the purpose of this article. We will 
extend the R ratio calculation in the MOM schemes of \cite{20,21} to the order 
beyond that computed in \cite{29} which will be to the same order of expansion 
as the R ratio in the $\MSbar$ and $\mMOM$ schemes. Therefore we will have a 
reasonable number of terms in the perturbative series to investigate whether 
the scheme dependence diminishes. We will carry this out not only for the R 
ratio but also for the flavour non-singlet Bjorken sum rule \cite{58,59}. One 
reason for considering this quantity as well is partly to see if a similar 
behaviour of scheme independence emerges. To quantify this in our approach in a
practical way we will gauge the theory error from each of the schemes at 
successive loop orders by extracting estimates for $\alpha_s(M_Z)$ using 
experimental data sets. It should be the case that at higher orders agreement 
on these ought to improve. We stress, however, that our study in the main will 
purely be in a theory laboratory. By this we mean quarks will be massless and 
we will ignore resonances that are inevitably present in experimental data. 
This is because we want to concentrate and particularly focus on the scheme 
dependence issue without the complication or distraction of other features. For
instance including quark masses in a MOM scheme R ratio evaluation is not 
straightforward. This is because to have analytic results for the three loop
R ratio would require two loop symmetric point $3$-point master integrals as a 
function of the quark masses in order to renormalize the Lagrangian in a MOM 
scheme as a function of mass. Such masters are not currently known. Once the 
scheme issues are understood in this idealized massless setting and the 
significance of the kinematic schemes quantified then quark mass effects could 
be included within our framework possibly via a numerical approach. Moreover 
modelling and including resonance effects for instance has not been tackled yet
for MOM schemes as such. It would necessitate the use of a controlled 
approximation dependent on an appropriate mass scale in each case.

The article is organized as follows. We provide an overview of the perturbative
expressions for the Bjorken sum rule and R ratio at high loop order in various
renormalization schemes in Section $2$. These then form the foundation for our
analysis of the scheme dependence in Section $3$ where for example we examine 
the convergence properties of an effective coupling constant derived from each
observable. The not unrelated issue of assigning an error to such an analysis
is discussed in Section $4$ for each of the schemes we consider and we also
provide estimates of $\alpha_s(M_Z)$ as a benchmark test. Finally concluding 
comments are provided in Section $5$. 

\sect{MOM scheme.}

Before discussing the construction of the MOM scheme expressions we recall the 
background to the two quantities of interest. First, the Bjorken sum rule 
originates from polarized deep inelastic scattering \cite{58,59} where it 
measures properties of the distribution of quark spins inside nucleons. In 
particular the integral over all momentum fractions is defined as
\begin{equation}
\Gamma_1^{p-n}(Q^2) ~=~ \int_0^1 \left[ g_1^{ep}(x,Q^2) ~-~ 
g_1^{en}(x,Q^2) \right] dx
\end{equation}
where $Q^2$~$=$~$-$~$q^2$ and $g_1^{ep}$ and $g_1^{en}$ are the structure 
functions associated with the spin of the proton ($p$) and neutron ($n$)
respectively. If the parton model point of view was valid in experiments then
$\Gamma_1^{p-n}(Q^2)$ would equate exactly to $\sixth g_A$ where $g_A$ is the
nucleon axial charge deduced from neutron $\beta$-decay. However due to quantum
corrections deriving from QCD and from a wealth of experimental evidence the
equality does not hold and theoretically $\Gamma_1^{p-n}(Q^2)$ depends on the
strong coupling constant $a$ taking the form  
\begin{equation}
\Gamma_1^{p-n}(Q^2) ~=~ \frac{g_A}{6} 
C_{{\mbox{\footnotesize{Bjr}}}}(a,\alpha) ~+~
\sum_{r=2}^\infty \frac{\mu^{p-n}_{2r} (Q^2)}{(Q^2)^{r-1}} 
\end{equation}
where Bjr denotes the Bjorken sum rule and the second term represents 
contributions from twists higher than $2$ reflecting non-perturbative 
contributions. By constrast $C_{{\mbox{\footnotesize{Bjr}}}}(a,\alpha)$ is 
perturbatively accessible and is determined from computing the correlation 
function of the axial vector current in the operator product expansion. In 
particular the expansion of $C_{{\mbox{\footnotesize{Bjr}}}}(a,\alpha)$ begins 
with unity. Our focus will be on this contribution and not the non-perturbative
piece. We have included the gauge parameter $\alpha$ in the argument of 
$C_{{\mbox{\footnotesize{Bjr}}}}(a,\alpha)$ since such dependence will in 
general be present in many schemes although it is absent in the $\MSbar$
scheme. 

The second quantity of interest is the hadronic R ratio which is related to
the correlation function of the electromagnetic current $j_\mu$. It is defined 
by
\begin{equation}
R(s) ~=~ 12 \pi \, \mbox{Im} \, \Pi( - s - i \varepsilon )
\end{equation}
with $\Pi(Q^2)$ connected to the correlation function via
\begin{equation}
\Pi_{\mu\nu}(Q^2) ~=~ \frac{i}{Q^2} \int \, d^4x \, e^{iqx} 
\langle 0| T j_\mu(x) j_\nu(0) |0 \rangle
\end{equation}
where
\begin{equation}
\Pi_{\mu\nu}(q^2) ~=~ -~ \left[ q^2 \eta_{\mu\nu} - q_\mu q_\nu \right]
\Pi(q^2) ~.
\end{equation}
For practical purposes we remove a common factor by defining
\begin{equation}
R(s) ~=~ \NF \left( \sum_f Q_f^2 \right) r(a(s))
\end{equation}
where $\NF$ is the dimension of the fundamental representation of the colour
group and $Q_f$ is the charge of the active quark flavours at energy $s$. We
note that when the correlation function contains flavour singlet currents
$r(a(s))$ will involve terms involving another function of $Q_f$. Our focus
here however will be on the non-singlet case. Singlet terms begin at $O(a^3)$. 

Having introduced the background to the Bjorken sum rule and R ratio we now
discuss the derivation of the $O(a^4)$ loop expressions for each in the various
kinematic schemes associated with momentum subtraction. To construct the 
respective MOM scheme results at $O(a^4)$ we exploit a property of the 
renormalization group equation. Roughly stated if a quantity is available at
$\mbox{\L}$ loops in one scheme then it can be deduced at the same loop order 
in another scheme from the relation of the coupling constant in the new scheme 
to that in the old scheme at $(\mbox{\L}-1)$ loops. To achieve this requires the
$\beta$-function at $(\mbox{\L}-1)$ loops in the new scheme. These have been 
provided recently for the three MOM schemes in \cite{24} in the Landau gauge,
defined by $\alpha$~$=$~$0$, where QCD was renormalized to three loops in each 
of the three MOM schemes extending the one and two loop computations of 
\cite{20,21,23}. Therefore we have used the coupling constant maps recorded in 
\cite{20,21,23,24} and applied them to the two objects of interest in the 
Landau gauge. While the two loop maps have been constructed for arbitrary 
$\alpha$ the three loop MOM $\beta$-function computation was only carried out
solely for the Landau gauge, \cite{24}. This is sufficient for our analysis. As
the full expressions are long, in order to illustrate the structure of what 
emerges we record the $\MOMq$ expressions both for the Bjorken sum rule and the
R ratio in $SU(3)$. An important tool employed in this respect in our 
calculations throughout was the symbolic manipulation language {\sc Form}, 
\cite{60,61}. In the case of the Bjorken sum rule we have 
\begin{eqnarray}
\left. C^{\MOMqs}_{{\mbox{\footnotesize{Bjr}}}}(a,0) \right|_{SU(3)} &=& 
1 - 4 a ~+~ [ 72 \Nf - 510 \psid + 340 \pi^2 + 1269 ] \frac{a^2}{81}
\nonumber \\
&&
+~ [ 1258848 \psid \Nf 
- 233280 \Nf^2 
- 11664 \psiddd \Nf 
+ 31104 \pi^4 \Nf 
\nonumber \\
&& ~~~~
- 839232 \pi^2 \Nf 
+ 3094848 \zeta_3 \Nf 
- 5598720 \zeta_5 \Nf 
+ 6114528 \Nf 
\nonumber \\
&& ~~~~
- 962190 \psid^2 
+ 1282920 \psid \pi^2 
- 8916966 \psid 
\nonumber \\
&& ~~~~
- 13635 \psiddd 
- 391280 \pi^4 
+ 5944644 \pi^2 
- 21121074 \zeta_3 
\nonumber \\
&& ~~~~
+ 92378880 \zeta_5 
- 52764048 ] \frac{a^3}{52488} 
\nonumber \\
&&
+~ [ 20623196160 H_5 \Nf 
+ 187812172800 H_5 
+ 2267136 H_6 
\nonumber \\
&& ~~~~
+ 9405849600 \Nf^3 
- 2821754880 \psid^2 \Nf^2 
\nonumber \\
&& ~~~~
+ 3762339840 \psid \pi^2 \Nf^2 
- 15049359360 \psid \zeta_3 \Nf^2 
\nonumber \\
&& ~~~~
- 169462056960 \psid \Nf^2 
+ 1516838400 \psiddd \Nf^2 
\nonumber \\
&& ~~~~
- 5299015680 \pi^4 \Nf^2 
+ 10032906240 \pi^2 \zeta_3 \Nf^2 
\nonumber \\
&& ~~~~
+ 112974704640 \pi^2 \Nf^2 
- 135444234240 \zeta_3^2 \Nf^2 
\nonumber \\
&& ~~~~
+ 750691307520 \zeta_3 \Nf^2 
- 790091366400 \zeta_5 \Nf^2 
\nonumber \\
&& ~~~~
- 366253332480 \Nf^2 
+ 40828354560 \psid^3 \Nf  
\nonumber \\
&& ~~~~
- 81656709120 \psid^2 \pi^2 \Nf 
+ 155496983040 \psid^2 \Nf 
\nonumber \\
&& ~~~~
- 4996857600 \psid \psiddd \Nf 
+ 67762759680 \psid \pi^4 \Nf 
\nonumber \\
&& ~~~~
- 207329310720 \psid \pi^2 \Nf 
+ 1134305399808 \psid \zeta_3 \Nf 
\nonumber \\
&& ~~~~
- 1598994432000 \psid \zeta_5 \Nf 
+ 6622411146624 \psid \Nf 
\nonumber \\
&& ~~~~
+ 3331238400 \psiddd \pi^2 \Nf 
- 147525507360 \psiddd \Nf 
\nonumber \\
&& ~~~~
+ 452620224 \psiddddd \Nf 
- 41901705216 \pi^6 \Nf 
+ 462511123200 \pi^4 \Nf 
\nonumber \\
&& ~~~~
- 756203599872 \pi^2 \zeta_3 \Nf 
+ 1065996288000 \pi^2 \zeta_5 \Nf 
\nonumber \\
&& ~~~~
- 4414940764416 \pi^2 \Nf 
+ 4469659729920 \zeta_3^2 \Nf 
\nonumber \\
&& ~~~~
- 12921382559232 \zeta_3 \Nf 
+ 7683712081920 \zeta_5 \Nf 
\nonumber \\
&& ~~~~
+ 8295959347200 \zeta_7 \Nf 
- 318923619840 \Nf 
- 453980630880 \psid^3 
\nonumber \\
&& ~~~~
+ 907961261760 \psid^2 \pi^2 
- 2555163443700 \psid^2 
\nonumber \\
&& ~~~~
- 11132857680 \psid \psiddd 
- 575619887360 \psid \pi^4 
\nonumber \\
&& ~~~~
+ 3406884591600 \psid \pi^2 
+ 1202060938176 \psid \zeta_3 
\nonumber \\
&& ~~~~
+ 26383408128000 \psid \zeta_5 
- 41716811068632 \psid 
\nonumber \\
&& ~~~~
+ 7421905120 \psiddd \pi^2 
+ 1041696774405 \psiddd 
\nonumber \\
&& ~~~~
- 2561413512 \psiddddd 
+ 233115257088 \pi^6 
- 3913486262280 \pi^4 
\nonumber \\
&& ~~~~
- 801373958784 \pi^2 \zeta_3 
- 17588938752000 \pi^2 \zeta_5 
+ 27811207379088 \pi^2 
\nonumber \\
&& ~~~~
- 36874692771840 \zeta_3^2 
+ 97876178446536 \zeta_3 
- 48763797737760 \zeta_5 
\nonumber \\
&& ~~~~
- 136883329228800 \zeta_7 
+ 44612309619360 ] \frac{a^4}{3174474240} 
\nonumber \\
&& ~~~~
+~ O(a^5)
\label{bjrkmomqlan}
\end{eqnarray}
where the variable $a$ here is in the $\MOMq$ scheme. We use the convention 
that when an object is labelled by a scheme then the variables of the 
expression are in that scheme. We follow the notation of \cite{24} which 
results in a variety of numbers such as the Riemann zeta series $\zeta_n$ and 
the odd derivatives of the Euler $\psi$-function, $\psi(z)$, up to the fifth 
order. The first and third order derivatives appeared first in the one and two 
loop expressions respectively. However comparing the form of the two loop term 
of the coupling constant mapping for the MOM schemes with the same terms given 
in \cite{24} it is clear that the latter expressions are much simpler. This was
highlighted in \cite{24} and is due to a relation between harmonic 
polylogarithms that were present in the two loop symmetric point master 
integrals computed directly in \cite{62,63,64,65,66}. In \cite{24,67} an 
indirect method was chosen to compute the masters. This involved using the 
{\sc Mincer} algorithm, \cite{26}, to evaluate the symmetric point masters by 
using a Taylor series expansion in the limit as one of the external momenta 
approaches zero, \cite{25}. The resulting expression was computed to very high 
numerical accuracy and then reconstructed analytically using the {\tt PSLQ} 
algorithm, \cite{27}. At two loops the basis choice used for the {\tt PSLQ} fit
included rationals, $\pi$, $\zeta_2$, $\zeta_3$, $\zeta_4$, $\zeta_5$, $\psid$ 
and $\psiddd$ and combinations. The absence of harmonic polylogarithms that 
appeared in previous work, \cite{23}, is because it is now known that they were 
not independent of the basis choice \cite{24}. One specific relation was 
recorded in \cite{68} which we recall is
\begin{eqnarray}
s_2 \left( \pisix \right) &=& \frac{1}{11664} 
\left[ 324 \sqrt{3} \ln(3) \pi - 27 \sqrt{3} \ln^2(3) \pi + 29 \sqrt{3} \pi^3 
- 1944 \psid 
\right. \nonumber \\
&& \left. ~~~~~~~~~ 
+ 23328 s_2 \left( \pitwo \right) + 19440 s_3 \left( \pisix \right) 
- 15552 s_3 \left( \pitwo \right) + 1296 \pi^2 \right]
\label{harmpolrel}
\end{eqnarray}
and holds numerically where
\begin{equation}
s_n(z) ~=~ \frac{1}{\sqrt{3}} \mbox{Im} \, \left[ \mbox{Li}_n \left(
\frac{e^{iz}}{\sqrt{3}} \right) \right]
\end{equation}
and $\mbox{Li}_n(z)$ is the polylogarithm function. The linear combination of 
$s_2 \left( \pitwo \right)$, $s_2 \left( \pisix \right)$,
$s_3 \left( \pitwo \right)$ and $s_3 \left( \pisix \right)$ that appear in the 
two loop masters of \cite{62,63,64,65,66} together with the terms with an odd 
power of $\pi$ is the same as that which occurs in (\ref{harmpolrel}). At 
$O(a^4)$ the quantities $H_5$ and $H_6$ appear and the lengthy definitions for 
them are given explicitly in terms of both generalized and harmonic 
polylogarithms in the {\tt info.pdf} file accessible from the Supplemental 
Material associated with \cite{24}. The three loop master integrals that 
underpin \cite{24} are recorded in \cite{67}. We note that electronic versions 
of (\ref{bjrkmomqlan}) as well as similar expressions in the other Landau gauge
MOM schemes are available for an arbitrary colour group in the data file of the
associated arXiv version of this article. It also contains the parallel data 
for the R ratio.
 
While the expressions for the other two kinematic schemes have a similar
structure what is perhaps more instructive is to record the numerical 
expressions for all the schemes we will consider here. For instance for the
$SU(3)$ group we have
\begin{eqnarray}
\left. C^{\MSbars}_{{\mbox{\footnotesize{Bjr}}}}(a,0) \right|_{SU(3)} &=& 
1 - 4.000000 a + [ 5.333333 \Nf - 73.333333 ] a^2 \nonumber \\
&& +~ [ - 11.358025 \Nf^2 + 486.866536 \Nf - 2652.1544368 ] a^3 \nonumber \\
&& +~ [ 26.556927 \Nf^3 - 1970.551732 \Nf^2 + 31588.209324 \Nf
\nonumber \\
&& ~~~~
- 122738.570412 ] a^4 ~+~ O(a^5)
\nonumber \\
\left. C^{\MOMgs}_{{\mbox{\footnotesize{Bjr}}}}(a,0) \right|_{SU(3)} &=& 
1 - 4.000000 a + [ - 8.333892 \Nf + 32.636622 ] a^2 \nonumber \\
&& +~ [ - 37.559047 \Nf^2 + 343.123637 \Nf - 539.554265 ] a^3 \nonumber \\
&& +~ [ - 190.378849 \Nf^3 + 2382.689181 \Nf^2 - 8789.557343 \Nf 
\nonumber \\
&& ~~~~
+ 3685.269188 ] a^4 \nonumber ~+~ O(a^5)
\nonumber \\
\left. C^{\MOMqs}_{{\mbox{\footnotesize{Bjr}}}}(a,0) \right|_{SU(3)} &=& 
1 - 4.000000 a + [ 0.888889 \Nf - 6.470235 ] a^2 \nonumber \\
&& +~ [ - 4.444444 \Nf^2 + 110.326436 \Nf - 547.208073 ] a^3 \nonumber \\
&& +~ [ 2.962963 \Nf^3 - 260.408043 \Nf^2 + 4464.055445 \Nf 
- 19390.815675 ] a^4 \nonumber \\
&& +~ O(a^5)
\nonumber \\
\left. C^{\MOMcs}_{{\mbox{\footnotesize{Bjr}}}}(a,0) \right|_{SU(3)} &=& 
1 - 4.000000 a + [ 0.888889 \Nf + 0.8597681 ] a^2 \nonumber \\
&& +~ [ - 4.444444 \Nf^2 + 113.578619 \Nf - 116.288597 ] a^3 \nonumber \\
&& +~ [ 2.962963 \Nf^3 - 195.703375 \Nf^2 + 2213.334884 \Nf 
- 10426.816777 ] a^4 \nonumber \\
&& +~ O(a^5)
\nonumber \\
\left. C^{\mMOMs}_{{\mbox{\footnotesize{Bjr}}}}(a,0) \right|_{SU(3)} &=& 
1 - 4.000000 a + [ 0.888889 \Nf - 17.000000 ] a^2 \nonumber \\
&& +~ [ - 4.444444 \Nf^2 + 134.233344 \Nf - 271.420979 ] a^3 \nonumber \\
&& +~ [ 2.962963 \Nf^3 - 278.134393 \Nf^2 + 4556.133051 \Nf 
- 17123.525160 ] a^4 \nonumber \\
&& +~ O(a^5)
\label{bjrknum}
\end{eqnarray}
in the Landau gauge where the $O(a^4)$ $\MSbar$ result was provided in \cite{9}
and that for the minimal momentum ($\mMOM$) scheme was given in \cite{69}. 
Clearly the $O(a)$ term is scheme independent. The higher order terms show no 
commonality. For instance the sign of the leading $\Nf$ term at each order 
varies from scheme to scheme. Moreover the $O(a^4)$ $\Nf$ independent term 
ranges over two orders of magnitude from the $\MSbar$ scheme to the $\MOMg$ 
one. This of course does not mean that 
$C^{\MSbars}_{{\mbox{\footnotesize{Bjr}}}}(a,0)$ is not as accurate as the 
expression in the latter scheme. The reason for this is that aside from the 
fact that the expansion variables are different in each scheme, the coupling
constant is itself a function that depends on the underlying momentum scale. 
Explicitly including the momentum dependence of the running coupling in each 
scheme balances out the apparent disparity in the coefficients we mentioned.

Having focused on the derivation of the Bjorken sum rule in kinematic schemes
we have repeated the exercise for the R ratio. So for instance the $\MOMq$
scheme version is 
\begin{eqnarray}
\left. r^{\MOMqs}(a) \right|_{SU(3)} &=&
1 + 4 a 
+ [ 864 \zeta_3 \Nf - 828 \Nf + 510 \psid - 340 \pi^2 - 14256 \zeta_3 + 12501 ]
\frac{a^2}{81} 
\nonumber \\
&& 
+ \left[
\left[ 855360 - 1866240 \zeta_3 \right] \eta 
- 31104 \pi^2 \Nf^2 
- 1119744 \zeta_3 \Nf^2 
+ 1679616 \Nf^2 
\right. \nonumber \\
&& \left. ~~~~
+ 1762560 \psid \zeta_3 \Nf 
- 2801088 \psid \Nf 
+ 11664 \psiddd \Nf
\right. \nonumber \\
&& \left. ~~~~
- 31104 \pi^4 \Nf
- 1175040 \pi^2 \zeta_3 \Nf
+ 2893824 \pi^2 \Nf
+ 53047872 \zeta_3 \Nf
\right. \nonumber \\
&& \left. ~~~~
- 9331200 \zeta_5 \Nf
- 59723568 \Nf 
+ 962190 \psid^2
- 1282920 \psid \pi^2
\right. \nonumber \\
&& \left. ~~~~
- 29082240 \psid \zeta_3 
+ 37007766 \psid 
+ 13635 \psiddd 
+ 391280 \pi^4 
\right. \nonumber \\
&& \left. ~~~~
+ 19388160 \pi^2 \zeta_3
- 33139908 \pi^2
- 510524046 \zeta_3 
+ 153964800 \zeta_5 
\right. \nonumber \\
&& \left. ~~~~
+ 455543352 \right] \frac{a^3}{52488} 
\nonumber \\
&& 
+ \left[ 
\left[ 225740390400 \zeta_3^2 \Nf 
+ 357422284800 \zeta_3 \Nf
- 564350976000 \zeta_5 \Nf
\right. \right. \nonumber \\
&& \left. \left. ~~~~
- 294716620800 \Nf
- 532998144000 \psid \zeta_3 
\right. \right. \nonumber \\
&& \left. \left. ~~~~
+ 244290816000 \psid 
+ 355332096000 \pi^2 \zeta_3
- 162860544000 \pi^2
\right. \right. \nonumber \\
&& \left. \left. ~~~~
- 3724716441600 \zeta_3^2
- 7976160460800 \zeta_3 
+ 10628610048000 \zeta_5 
\right. \right. \nonumber \\
&& \left. \left. ~~~~
+ 4722912230400 \right] \eta
\right. \nonumber \\
&& \left. ~~~~
- 20623196160 H_5 \Nf
- 187812172800 H_5
- 2267136 H_6 
\right. \nonumber \\
&& \left. ~~~~
- 15049359360 \pi^2 \zeta_3 \Nf^3
+ 14422302720 \pi^2 \Nf^3
+ 158018273280 \zeta_3 \Nf^3
\right. \nonumber \\
&& \left. ~~~~
+ 225740390400 \zeta_5 \Nf^3
- 378115153920 \Nf^3 
+ 2821754880 \psid^2 \Nf^2
\right. \nonumber \\
&& \left. ~~~~
- 12645642240 \psid \pi^2 \Nf^2
- 663425925120 \psid \zeta_3 \Nf^2
\right. \nonumber \\
&& \left. ~~~~
+ 896377466880 \psid \Nf^2
+ 3762339840 \psiddd \zeta_3 \Nf^2
\right. \nonumber \\
&& \left. ~~~~
- 4808885760 \psiddd \Nf^2
- 10032906240 \pi^4 \zeta_3 \Nf^2
+ 20000010240 \pi^4 \Nf^2
\right. \nonumber \\
&& \left. ~~~~
+ 1187227238400 \pi^2 \zeta_3 \Nf^2
- 1380621957120 \pi^2 \Nf^2
\right. \nonumber \\
&& \left. ~~~~
+ 902961561600 \zeta_3^2 \Nf^2
- 12819963985920 \zeta_3 \Nf^2
\right. \nonumber \\
&& \left. ~~~~
- 9706836787200 \zeta_5 \Nf^2
+ 22470692267520 \Nf^2 
\right. \nonumber \\
&& \left. ~~~~
- 40828354560 \psid^3 \Nf
+ 81656709120 \psid^2 \pi^2 \Nf
\right. \nonumber \\
&& \left. ~~~~
+ 394261862400 \psid^2 \zeta_3 \Nf
- 500476112640 \psid^2 \Nf
\right. \nonumber \\
&& \left. ~~~~
+ 4996857600 \psid \psiddd \Nf
- 67762759680 \psid \pi^4 \Nf
\right. \nonumber \\
&& \left. ~~~~
- 525682483200 \psid \pi^2 \zeta_3 \Nf
+ 960450462720 \psid \pi^2 \Nf
\right. \nonumber \\
&& \left. ~~~~
+ 24529502979072 \psid \zeta_3 \Nf
- 2664990720000 \psid \zeta_5 \Nf
\right. \nonumber \\
&& \left. ~~~~
- 30896934171264 \psid \Nf
- 3331238400 \psiddd \pi^2 \Nf
\right. \nonumber \\
&& \left. ~~~~
- 57680501760 \psiddd \zeta_3 \Nf
+ 203639456160 \psiddd \Nf
\right. \nonumber \\
&& \left. ~~~~
- 452620224 \psiddddd \Nf
+ 41901705216 \pi^6 \Nf
+ 329042165760 \pi^4 \zeta_3 \Nf
\right. \nonumber \\
&& \left. ~~~~
- 960905030400 \pi^4 \Nf
- 28644566243328 \pi^2 \zeta_3 \Nf
\right. \nonumber \\
&& \left. ~~~~
+ 1776660480000 \pi^2 \zeta_5 \Nf
+ 33903941250816 \pi^2 \Nf
\right. \nonumber \\
&& \left. ~~~~
- 46905874606080 \zeta_3^2 \Nf
+ 280279032268032 \zeta_3 \Nf
\right. \nonumber \\
&& \left. ~~~~
+ 124706756321280 \zeta_5 \Nf
+ 7505867980800 \zeta_7 \Nf
\right. \nonumber \\
&& \left. ~~~~
- 389487227794560 \Nf 
+ 453980630880 \psid^3 
\right. \nonumber \\
&& \left. ~~~~
- 907961261760 \psid^2 \pi^2 
- 6505320729600 \psid^2 \zeta_3 
\right. \nonumber \\
&& \left. ~~~~
+ 8838711875700 \psid^2 
+ 11132857680 \psid \psiddd 
\right. \nonumber \\
&& \left. ~~~~
+ 575619887360 \psid \pi^4 
+ 8673760972800 \psid \pi^2 \zeta_3 
\right. \nonumber \\
&& \left. ~~~~
- 14203428246000 \psid \pi^2 
- 214276142356416 \psid \zeta_3 
\right. \nonumber \\
&& \left. ~~~~
+ 43972346880000 \psid \zeta_5 
+ 215899216511832 \psid 
\right. \nonumber \\
&& \left. ~~~~
- 7421905120 \psiddd \pi^2 
- 72568742400 \psiddd \zeta_3 
\right. \nonumber \\
&& \left. ~~~~
- 971601966405 \psiddd 
+ 2561413512 \psiddddd 
- 233115257088 \pi^6 
\right. \nonumber \\
&& \left. ~~~~
- 2697737011200 \pi^4 \zeta_3 
+ 8131574351880 \pi^4 
+ 210454364985984 \pi^2 \zeta_3 
\right. \nonumber \\
&& \left. ~~~~
- 29314897920000 \pi^2 \zeta_5 
- 215086534781328 \pi^2 
+ 528115645854720 \zeta_3^2 
\right. \nonumber \\
&& \left. ~~~~
- 1807448276772936 \zeta_3 
- 436869622959840 \zeta_5 
- 123846821683200 \zeta_7 
\right. \nonumber \\
&& \left. ~~~~
+ 2018890127348640 \right] \frac{a^4}{3174474240} ~+~ O(a^5)
\end{eqnarray}
for $SU(3)$, also in the Landau gauge, which includes a parameter $\eta$. It is
defined to be
\begin{equation}
\eta ~=~
\frac{ \sum_f Q_f^2 }{ \left[ \sum_f Q_f \right]^2 }
\end{equation}
and is associated with $O(a^3)$ graphs that contribute to the flavour singlet 
value of the R ratio. Similar to the Bjorken case we will set $\eta$~$=$~$0$ 
throughout and concentrate on the non-singlet scenario. Moreover, we have 
checked that all the MOM scheme $O(a^3)$ R ratio expressions given in \cite{29}
agree with the analytic $O(a^3)$ expressions derived using the coupling 
constant mappings provided in \cite{24} after taking (\ref{harmpolrel}) into 
account. For completeness we record the $O(a^4)$ numerical expressions for the 
R ratio in each scheme. We have 
\begin{eqnarray}
\left. r^{\MSbars}(a) \right|_{SU(3)} &=&
1 + 4.000000 a + [ - 1.844726 \Nf + 31.771318 ] a^2 \nonumber \\
&& +~ [ - 26.443505 \eta - 0.331415 \Nf^2 - 76.808579 \Nf - 424.763877 ] a^3 
\nonumber \\
&& +~ [ 49.056846 \eta \Nf - 1521.214892 \eta + 5.508123 \Nf^3 
- 204.143191 \Nf^2 \nonumber \\
&& ~~~~+~ 4806.339848 \Nf - 40091.676394 ] a^4
~+~ O(a^5) \nonumber \\
\left. r^{\MOMgs}(a) \right|_{SU(3)} &=&
1 + 4.000000 a + [ 11.822499 \Nf - 74.198637 ] a^2 \nonumber \\
&& +~ [ - 26.443505 \eta + 49.709397 \Nf^2 - 401.928163 \Nf 
- 335.201612 ] a^3 \nonumber \\
&& +~ [ - 222.000165 \eta \Nf + 580.4478702 \eta + 252.625523 \Nf^3
- 2960.042900 \Nf^2 \nonumber \\
&& ~~~~+~ 7418.803963 \Nf + 12015.778615 ] a^4
~+~ O(a^5) \nonumber \\
\left. r^{\MOMqs}(a) \right|_{SU(3)} &=&
1 + 4.000000 a + [ 2.599718 \Nf - 35.091780 ] a^2 \nonumber \\
&& +~ [ - 26.443505 \eta + 0.507465 \Nf^2 + 90.741952 \Nf 
- 1140.227696 ] a^3 \nonumber \\
&& +~ [ - 39.088169 \eta \Nf - 195.143901 \eta + 3.058056 \Nf^3 
- 183.223433 \Nf^2 \nonumber \\
&& ~~~~+~ 3501.125982 \Nf - 7958.138070 ] a^4
~+~ O(a^5) \nonumber \\
\left. r^{\MOMcs}(a) \right|_{SU(3)} &=&
1 + 4.000000 a + [ 2.599718 \Nf - 42.421783 ] a^2 \nonumber \\
&& +~ [ - 26.443505 \eta + 0.507465 \Nf^2 + 74.704019 \Nf 
- 1418.822322 ] a^3 \nonumber \\
&& +~ [ - 39.088169 \eta \Nf - 49.770672 \eta + 3.058056 \Nf^3 
- 237.639896 \Nf^2 \nonumber \\
&& ~~~~+~ 4083.180193 \Nf + 677.129492 ] a^4
~+~ O(a^5) \nonumber \\
\left. r^{\mMOMs}(a) \right|_{SU(3)} &=&
1 + 4.000000 a + [ 2.599718 \Nf - 24.562015 ] a^2 \nonumber \\
&& +~ [ - 26.443505 \eta + 0.507465 \Nf^2 + 85.202150 \Nf 
- 1634.833914 ] a^3 \nonumber \\
&& +~ [ - 39.088169 \eta \Nf - 403.976819 \eta + 3.058056 \Nf^3 
- 230.126428 \Nf^2 \nonumber \\
&& ~~~~+~ 4880.206237 \Nf - 17400.630113 ] a^4 ~+~ O(a^5) ~.
\label{rratnum}
\end{eqnarray}
Similar comments to the different scheme expressions of the Bjorken case apply
to the R ratio in that there is a large disparity in the numerical values. For
instance not only does the size of the coefficients of the $O(a^4)$ 
$\Nf$~$=$~$0$ terms have a wide spread but they can have either sign. 

\sect{Analysis.} \label{sec:Analysis}

Our analysis will involve the Bjorken sum rule and the R ratio evaluated in
the $\mMOM$, $\MSbar$ and the three MOM schemes of \cite{20,21}, that we will
collectively denote by $\MOMi$, to provide a comparison of measureable 
perturbative series considered in different renormalization schemes. As noted 
the numerical values in (\ref{bjrknum}) and (\ref{rratnum}) cannot be directly 
compared as they are written in terms of the coupling constants for different 
schemes. Therefore we must have a standard which we use to relate the coupling 
constants to one another. For example, one could choose to insert the coupling 
constant conversion functions without truncation to get the series in one 
scheme in terms of the coupling constant of another. However, as this only 
makes a connection with the value of an abstract formal quantity, we will 
instead choose to compare the series at set energy levels by substituting the 
explicit form of the coupling constant in terms of the unphysical 
renormalization momentum.

The expression for the coupling constant at a particular energy level can 
be found by solving the renormalization group equations perturbatively. 
Denoting the expression for the coupling constant to the \L-th loop level these
expressions are given by
\begin{eqnarray}
a_1^{\cal S}(Q,\Lambda^{\cal S}) &=&\label{eq:CC1L}
\frac{1}{b_0^{\cal S} L^{\cal S}}
, \nonumber \\
a_2^{\cal S}(Q,\Lambda^{\cal S}) &=&
\frac{1}{b_0^{\cal S} L^{\cal S}}
\left[ 1 - \frac{b_1^{\cal S} \ln (L^{\cal S})}{{b_0^{\cal S}}^2 L^{\cal S}}
\right], \nonumber \\
a_3^{\cal S}(Q,\Lambda^{\cal S}) &=&
\frac{1}{b_0^{\cal S} L^{\cal S}}
\left[ 1 - \frac{b_1^{\cal S} \ln (L^{\cal S})}{{b_0^{\cal S}}^2 L^{\cal S}}
+ \left[ {b_1^{\cal S}}^2 \left[ \ln^2 (L^{\cal S}) - \ln (L^{\cal S}) - 1 
\right]
+ b_0^{\cal S} b_2^{\cal S} \right] \frac{1}{{b_0^{\cal S}}^4 {L^{\cal S}}^2}
\right], \nonumber \\
a_4^{\cal S}(Q,\Lambda^{\cal S}) &=&
\frac{1}{b_0^{\cal S} L^{\cal S}}
\left[ 1 - \frac{b_1^{\cal S} \ln (L^{\cal S})}{{b_0^{\cal S}}^2 L^{\cal S}}
+ \left[ {b_1^{\cal S}}^2 \left[ \ln^2 (L^{\cal S}) - \ln (L^{\cal S}) - 1 
\right]
+ b_0^{\cal S} b_2^{\cal S} \right] \frac{1}{{b_0^{\cal S}}^4 {L^{\cal S}}^2}
\right. \nonumber \\
&& \left. ~~~~~~~~~
- \left[ {b_1^{\cal S}}^3 \left[ \ln^3 (L^{\cal S}) - \frac{5}{2} \ln^2 
(L^{\cal S})
- 2 \ln (L^{\cal S}) + \frac{1}{2} \right]
+ 3 b_0^{\cal S} b_1^{\cal S} b_2^{\cal S} \ln (L^{\cal S}) 
\right. \right. \nonumber \\
&& \left. \left. ~~~~~~~~~~~~~
- \frac{1}{2} {b_0^{\cal S}}^2 b_3^{\cal S} \right]
\frac{1}{{b_0^{\cal S}}^6 {L^{\cal S}}^3} \right], \nonumber \\ 
a_5^{\cal S}(Q,\Lambda^{\cal S}) &=&
\frac{1}{b_0^{\cal S} L^{\cal S}}
\left[ 1 - \frac{b_1^{\cal S} \ln (L^{\cal S})}{{b_0^{\cal S}}^2 L^{\cal S}}
+ \left[ {b_1^{\cal S}}^2 \left[ \ln^2 (L^{\cal S}) - \ln (L^{\cal S}) - 1 
\right]
+ b_0^{\cal S} b_2^{\cal S} \right] \frac{1}{{b_0^{\cal S}}^4 {L^{\cal S}}^2}
\right. \nonumber \\
&& \left. ~~~~~~~~~
- \left[ {b_1^{\cal S}}^3 \left[ \ln^3 (L^{\cal S}) - \frac{5}{2} \ln^2 
(L^{\cal S})
- 2 \ln (L^{\cal S}) + \frac{1}{2} \right]
+ 3 {b_0^{\cal S}} {b_1^{\cal S}} {b_2^{\cal S}} \ln (L^{\cal S}) 
\right. \right. \nonumber \\
&& \left. \left. ~~~~~~~~~~~~~
- \frac{1}{2} {b_0^{\cal S}}^2 b_3^{\cal S} \right]
\frac{1}{{b_0^{\cal S}}^6 {L^{\cal S}}^3} \right. \nonumber \\
&& \left. ~~~~~~~~~ + \left[18 {b_0^{\cal S}} {b_2^{\cal S}} {b_1^{\cal S}}^2 \left[2\ln^2(L^{\cal S})-\ln(L^{\cal S})-1\right] + 2 {b_0^{\cal S}}^2 \left[5 {{b_2^{\cal S}}}^2+ {b_0^{\cal S}} {b_4^{\cal S}}\right] \right. \right. \nonumber \\
&& \left.\left.~~~~~~~~~~~~~ + {b_1^{\cal S}}^4 \left[6 \ln^4(L^{\cal S})-26 \ln^3(L^{\cal S})-9\ln^2(L^{\cal S})+24 \ln(L^{\cal S})+7\right]\right. \right. \nonumber \\
&&  \left.\left.~~~~~~~~~~~~~  -{b_0^{\cal S}}^2 {b_3^{\cal S}} {b_1^{\cal S}} \left[12 \ln(L^{\cal S})+1\right] \right] \frac{1}{6 {b_0^{\cal S}}^6 {L^{\cal S}}^4}\label{eq:CC5L}
\right].
\label{eq:explicitcouplingconstant}
\end{eqnarray}
where ${\cal S}$ indicates the scheme,
$L^{\cal S}$~$=$~$\ln \left( \frac{Q^2}{{\Lambda^{\cal S}}^2} \right)$, the 
subscript on $a_{\mbox{\footnotesize{\L}}}^{\cal S}(Q,\Lambda^{\cal S})$ 
denotes the loop order that the running coupling constant is approximated to 
and $b_n^{\cal S}$ are the coefficients of the $O(a^{n+2})$ term in the 
$\beta$-function in scheme ${\cal S}$ \cite{70}. Within each of these equations
the momentum always appears in the combination $x$~$=$~$|Q|/\Lambda^{\cal S}$. 
Therefore we can consider this quantity as the running momentum in units of the 
$\Lambda$-parameter, $\Lambda^{\cal S}$, of scheme ${\cal S}$. This parameter 
is introduced as a constant of integration when solving for the coupling 
constant and indicates the position of the Landau pole where the leading order 
term becomes divergent. Its value is dependent on the number of active fermions
which will be left implicit unless the value of $\Nf$ we are considering is 
unclear. The $\Lambda$-parameter is also scheme dependent. However its value in
one scheme can be found in terms of that of another using the $\Lambda$-ratio. 
It is given {\em exactly} from a simple one loop calculation. See, for 
instance, \cite{21},
\begin{equation}
\frac{\Lambda^{{\cal S}_1}}{\Lambda^{{\cal S}_2}} ~=~ 
\exp\left(\frac{c^{{\cal S}_1,{\cal S}_2}_1}{2 b^{{\cal S}_1}_0}\right)
\label{eq:lamratio}
\end{equation}
where $c^{{\cal S}_1,{\cal S}_2}_1$ is the coefficient of the leading order 
coupling constant conversion function between the two schemes ${\cal S}_1$ and
${\cal S}_2$. Since the leading order $\beta$-function coefficient is scheme 
independent, $c^{{\cal S}_1,{\cal S}_2}_1$ is the only scheme dependent 
quantity in the equation.  This relation allows one to use the
$\Lambda$-parameter in a single scheme as the only input parameter to the 
theory for any scheme required to make numerical calculations.   

\subsection{Partial sum analysis.}

In \cite{21} a table was constructed with values of the R ratio at $O(a)$ and 
$O(a^2)$ level for a variety of different energies and number of active quarks 
in both the $\MSbar$ and $\MOMq$ schemes. The momenta considered were 
$Q$~$=$~$3$ GeV with $\Nf$~$=$~$3$, $Q$~$=$~$5$ GeV with $\Nf$~$=$~$4$, 
$Q$~$=$~$20$ GeV with $\Nf$~$=$~$5$ and $Q$~$=$~$40$ GeV with $\Nf$~$=$~$6$. 
Choices of $\Lambda^{\MSbars}$~$=$~$500$ MeV and $\Lambda^{\MOMqs}$~$=$~$850$
MeV were made then for the analysis that produced Table III of \cite{13} with 
these numbers chosen to agree with values regarded as reliable at the time of
\cite{13} but have been superseded by subsequent measurements.

An extension of this table was produced in Tables II and III of \cite{29} where
the values of the partial sum 
\begin{equation}
a_{pq}^{\cal S} \left( Q,\Lambda^{\cal S}\right) ~=~ 
r_1 \sum_{n=1}^p a_{RR,n}^{\cal S}(s) 
\left( a_q^{\cal S}(Q,\Lambda^{\cal S}) \right)^n ~.
\end{equation}
were calculated where $s$ is the centre of mass energy and $RR$ denotes the R
ratio. The three $\MOMi$ schemes were considered up to $p$, $q$~$=$~$3$ while 
$\mMOM$ and $\MSbar$ could be calculated to $p$, $q$~$=$~$4$. The value 
$\Lambda^{\MSbars}$~$=$~$500$ MeV was again taken with the $\Lambda$-parameter 
in the other schemes then being deduced from the exact $\Lambda$-ratio 
formalism. Only the $Q$~$=$~$20$ GeV and $Q$~$=$~$40$ GeV values were produced 
because the lower energies did not fall within the perturbative range 
considered in \cite{29}. Those tables allowed for the quantification of both 
the difference in the series in each scheme as well as the convergence of the 
value of the series in single scheme as the loop order is increased.

With the calculation of the R ratio up to $O(a^4)$ in the momentum subtraction 
schemes \cite{24} we have revisited Table III of \cite{13} and extended it to 
include all available information. The outcome is presented in Table 
\ref{Tab:CGTableNf5} for $Q$~$=$~$20$ GeV and in Table \ref{Tab:CGTableNf6} for
$Q$~$=$~$40$ GeV. The five loop $\beta$-function of both the $\MSbar$ scheme,
\cite{1,2,3}, and $\mMOM$ scheme, \cite{32}, have allowed for the inclusion of 
the values in these schemes at the five loop  level in the coupling constant. 
We note that while the choices of the values of $\Lambda^{\MSbars}$ and $\Nf$ 
at a particular energy level are not accurate to current phenomenological
measurements we have decided to continue with this data as an extension of past 
results and to have a common ground for comparison.

\begin{table}
\begin{center}
\begin{tabular}{ |c|c||c|c|c|c|c||c| }
 \hline 
  \rule{0pt}{12pt} 
 $p$ & $q$  &$\MSbar$ &$\MOMg$&$\MOMc$&$\MOMq$&$\mMOM$& Average \\
  \hline 
 1 & 1  &$ 0.0707$ &$ 0.0848 $&$ 0.0918 $&$ 0.0881 $&$ 0.0833 $&$0.0837 \pm 0.0080$\\ 
 1 & 2  &$ 0.0581$ &$ 0.0683 $&$ 0.0733 $&$ 0.0707 $&$ 0.0672 $&$0.0675 \pm 0.0058$\\ 
 1 & 3  &$ 0.0592$ &$ 0.0700 $&$ 0.0753 $&$ 0.0681 $&$ 0.0696 $&$0.0684 \pm 0.0058$\\ 
 1 & 4  &$ 0.0593$ &$ 0.0701 $&$ 0.0764 $&$ 0.0715 $&$ 0.0698 $&$0.0694 \pm 0.0062$\\ 
 1 & 5  &$ 0.0592$ &$-$&$-$&$-$&$ 0.0693 $&$(0.0642 \pm 0.0071)$\\ 
 \hline 
 2 & 2  &$ 0.0629$ &$ 0.0639 $&$ 0.0634 $&$ 0.0638 $&$ 0.0640 $&$0.0636 \pm 0.0005$\\ 
 2 & 3  &$ 0.0641$ &$ 0.0653 $&$ 0.0649 $&$ 0.0617 $&$ 0.0661 $&$0.0644 \pm 0.0017$\\ 
 2 & 4  &$ 0.0643$ &$ 0.0655 $&$ 0.0657 $&$ 0.0645 $&$ 0.0663 $&$0.0652 \pm 0.0008$\\ 
 2 & 5  &$ 0.0642$ &$-$&$-$&$-$&$ 0.0658 $&$(0.0650 \pm 0.0011)$\\ 
 \hline 
 3 & 3  &$ 0.0615$ &$ 0.0594 $&$ 0.0580 $&$ 0.0584 $&$ 0.0598 $&$0.0594 \pm 0.0014$\\ 
 3 & 4  &$ 0.0616$ &$ 0.0596 $&$ 0.0585 $&$ 0.0606 $&$ 0.0599 $&$0.0600 \pm 0.0012$\\ 
 3 & 5  &$ 0.0615$ &$-$&$-$&$-$&$ 0.0596 $&$(0.0606 \pm 0.0014)$\\ 
 \hline 
 4 & 4  &$ 0.0606$ &$ 0.0602 $&$ 0.0605 $&$ 0.0612 $&$ 0.0601 $&$0.0605 \pm 0.0004$\\ 
 4 & 5  &$ 0.0605$ &$- $&$-$&$ - $&$ 0.0597 $&$(0.0601 \pm 0.0006)$\\ 
\hline 
 \end{tabular} 
 \end{center}\caption{{Values of $a_{pq}^{\cal S}(Q,\Lambda^{\cal S})$ for
$\Nf$~$=$~$5$ with $\Lambda^{\MSbars}$~$=$~$500$ MeV at $Q$~$=$~$20$ GeV. 
Average taken is arithmetic mean for values of each scheme with the standard 
deviation as the error, the bracketed averages are averages produced from an 
incomplete set of schemes.} }\label{Tab:CGTableNf5}
 \end{table}

\begin{table}
\begin{center}
\begin{tabular}{ |c|c||c|c|c|c|c||c| }
 \hline 
  \rule{0pt}{12pt} 
 $p$ & $q$  &$\MSbar$ &$\MOMg$&$\MOMc$&$\MOMq$&$\mMOM$& Average \\
  \hline 
 1 & 1  &$ 0.0652$ &$ 0.0723 $&$ 0.0809 $&$ 0.0780 $&$ 0.0742 $&$0.0741 \pm 0.0060$\\ 
 1 & 2  &$ 0.0566$ &$ 0.0622 $&$ 0.0690 $&$ 0.0667 $&$ 0.0637 $&$0.0637 \pm 0.0047$\\ 
 1 & 3  &$ 0.0569$ &$ 0.0617 $&$ 0.0688 $&$ 0.0634 $&$ 0.0641 $&$0.0630 \pm 0.0043$\\ 
 1 & 4  &$ 0.0571$ &$ 0.0619 $&$ 0.0698 $&$ 0.0656 $&$ 0.0645 $&$0.0638 \pm 0.0047$\\ 
 1 & 5  &$ 0.0570$ &$ - $&$ - $&$ - $&$ 0.0644 $&$(0.0607 \pm 0.0052)$\\ 
 \hline 
 2 & 2  &$ 0.0608$ &$ 0.0614 $&$ 0.0610 $&$ 0.0613 $&$ 0.0615 $&$0.0612 \pm 0.0003$\\ 
 2 & 3  &$ 0.0611$ &$ 0.0610 $&$ 0.0609 $&$ 0.0585 $&$ 0.0618 $&$0.0607 \pm 0.0012$\\ 
 2 & 4  &$ 0.0613$ &$ 0.0611 $&$ 0.0616 $&$ 0.0604 $&$ 0.0621 $&$0.0613 \pm 0.0006$\\ 
 2 & 5  &$ 0.0613$ &$ - $&$ - $&$ - $&$ 0.0621 $&$(0.0617 \pm 0.0006)$\\ 
 \hline 
 3 & 3  &$ 0.0585$ &$ 0.0574 $&$ 0.0560 $&$ 0.0562 $&$ 0.0572 $&$0.0571 \pm 0.0010$\\ 
 3 & 4  &$ 0.0587$ &$ 0.0575 $&$ 0.0566 $&$ 0.0578 $&$ 0.0575 $&$0.0576 \pm 0.0008$\\ 
 3 & 5  &$ 0.0586$ &$ - $&$ -$&$ - $&$ 0.0575 $&$(0.0581 \pm 0.0008)$\\ 
 \hline 
 4 & 4  &$ 0.0580$ &$ 0.0578 $&$ 0.0582 $&$ 0.0584 $&$ 0.0578 $&$0.0580 \pm 0.0002$\\ 
 4 & 5  &$ 0.0579$ &$ - $&$ - $&$ - $&$ 0.0577 $&$(0.0578 \pm 0.0001)$\\ 
\hline 
 \end{tabular} 

 \end{center}\caption{{Values of $a_{pq}^{\cal S}(Q,\Lambda^{\cal S})$ for
$\Nf$~$=$~$6$ with $\Lambda^{\MSbars}$~$=$~$500$ MeV at $Q$~$=$~$40$ GeV. 
Average taken is arithmetic mean for values of each scheme with the standard 
deviation as the error, the bracketed averages are averages produced from an 
incomplete set of schemes.} }\label{Tab:CGTableNf6}
 \end{table}

The tables are filled by substituting the energy levels and values of the
$\Lambda$-parameter into the partial sum $a_{pq}^{\cal S}$ employing the 
explicit coupling constant expansion $a_q^{\cal S}(Q,\Lambda^{\cal S})$. Using 
the $\Lambda$-ratio and a choice of $\Lambda^{\MSbars}$ we can then compare the
values calculated  at a set loop order but in different schemes (along a row) 
or consider the convergence of a single value (down a column). In addition to 
columns for each scheme, we have included a column that acts as an average of 
the series. We will use the arithmetic mean for this average 
\begin{eqnarray}
\bar{a}_{pq}(Q) ~=~ \frac{1}{|S|}\sum_{s_i\in S} a_{pq}^{s_i}(Q,\Lambda^{s_i})
\end{eqnarray}
where $S$ is the set of schemes and $|S|$ is the length of the set. For most
loop orders considered the set is $S$~$=$~$\{\MSbar,\MOMg,\MOMc,\MOMq,\mMOM\}$.
However with the inclusion of the five loop coupling constant in both $\MSbar$ 
and $\mMOM$ schemes we take $S$~$=$~$\{\MSbar, \mMOM\}$ for $q$~$=$~$5$. 
Averages found for $|S|<5$ will be bracketed to indicate that a reduced set of 
schemes are considered. The unbiased standard deviation can then be used as the
error and is defined as
\begin{eqnarray}
\Delta a_{pq}=\sqrt{\frac{1}{|S|-1}{\sum_{s_i\in S}
\Big[ a_{pq}^{s_i}(Q,\Lambda^{s_i})-\bar{a}_{pq}(Q)\Big]^2}} ~.
\label{unbsigma}
\end{eqnarray}
In doing this we have implicitly assumed that all schemes are independent and 
are distributed about the `true' value of the series. This assumption is not 
strictly accurate as will be discussed in Section \ref{sec:Scheme Error} but 
for the purposes of providing a {\em rough} estimate of scheme dependence we 
will ignore this caveat. An example of a scheme that is not independent in the 
context of (\ref{unbsigma}) is the modified regularization invariant ($\RI$) 
scheme of \cite{71,72}. While the renormalization of the underlying fields 
differ from that of the $\MSbar$ scheme the renormalization of the coupling 
constant is carried out in an $\MSbar$ fashion. This in turn means that the 
$\RI$ scheme $\beta$-function is formally the same as the $\MSbar$ one which 
implies that the expressions for the R ratio and Bjorken sum rule are also 
formally the same as their $\MSbar$ counterparts. This follows from the all 
orders relation of the coupling constants to each other which is 
$a^{\MSbars}$~$=$~$a^{\RIs}$. Therefore including the $\RI$ scheme in the set 
$S$ would unnecessarily introduce a bias in determining $\Delta a_{pq}$.

Comparing the values in different schemes at a set loop order elucidates
the residual variation in the evaluation of the series due to the truncation
of the perturbative series at that order. First we consider the values in Table
1 for the leading order series which is $p$~$=$~$1$. These are the coupling
constants in each scheme at a particular energy multiplied by a constant
characteristic of the series. While the bare coupling constant in the
Lagrangian is renormalization group invariant, the running coupling depends
on the choice of renormalization scheme in such a way that all measureables do 
not when calculated to all orders. This means we do not expect to see a great 
reduction in scheme dependence by simply increasing the order of the coupling 
constant itself without including further terms in the series. This is shown by
considering the range of values at $q$~$=$~$1$ of $(0.0707, 0.0918)$ which is 
commensurate with the range for $q$~$=$~$4$ of $(0.0593, 0.0764)$. Note the 
form of $p$~$=$~$q$~$=$~$1$ is the same in all schemes with the only 
differences being introduced as a result of the various $\Lambda$-parameters. 

The range of values for $p$~$=$~$2$ significantly decreases with an absolute 
range of $0.0011$ at $q$~$=$~$2$, $0.0044$ at $q$~$=$~$3$ and $0.0020$ at 
$q$~$=$~$1$. This is in contrast to the increase at $p$~$=$~$3$ with ranges of 
$0.0035$ for $q$~$=$~$3$ and $0.0031$ for $q$~$=$~$4$. Since the apparent 
reduced scheme dependence at the two loop level is not continued when the three
loop coupling constant is included this suggests that it is likely not to be
true scheme independence. The reduced range of $0.0011$ at $p$~$=$~$q$~$=$~$4$ 
may be due to the series settling down towards its true value. However the 
large amount of reduction in scheme dependence may again be anomalous as it was
for the two loop case. Considering the full scheme dependence at 
$p$~$=$~$q$~$=$~$5$ would allow for the discernment of these two cases. Next we
balance this behaviour in relation to the standard deviation quoted as the 
error on the average for each loop order. Again we see a large scheme 
difference at $p$~$=$~$1$ for all $q$ since these values represent only the 
scheme dependent coupling constant difference reduced severely for 
$p$~$=$~$q$~$=$~$2$ and oscillates at larger values for $p$~$=$~$2$, 
$q$~$=$~$3$ and $q$~$=$~$4$. The error value is similarly increased at 
$p$~$=$~$3$ with over double the error shown at two loops. Finally at 
$p$~$=$~$q$~$=$~$4$ the error is reduced below all the other errors given. The 
averages for the five loop $\mMOM$ and $\MSbar$ coupling constant values at
each order in the series are of similar size but see a slight increase in error
over the four loop values at each order in the series. At other values for 
$1$~$<$~$p$~$<$~$q$ we observe that the quoted error behaves qualitatively like 
the average for $p$~$=$~$q$, increasing for $q$~$=$~$3$ and decreasing at 
$q$~$=$~$4$. Therefore provided the assumed scheme distribution is accurate 
this suggests a small increase in apparent scheme dependence in the R ratio 
series at $p$~$=$~$q$~$=$~$5$ in the series we considered. However, this could
not be concretely verified without the full calculation.

We will now discuss the convergence of the values in each scheme as loop order 
is increased by comparing values down a column of the table and first consider 
the convergence for $p$~$=$~$q$. In general the $p$~$=$~$1$ and $p$~$=$~$2$ 
values in all schemes are larger than for the $p$~$=$~$3$ and $p$~$=$~$4$. As 
is expected for a perturbative expansion as the loop order is increased the 
difference between $a^s_{N+1~N+1}$ and $a^S_{NN}$ decreases. For example, for 
the $\MSbar$ scheme the differences are $0.0068$, $0.0024$ and $0.0009$. We see
that the different schemes do not all converge in the same way. While the
$\MSbar$ scheme decreases at each order, converging from above, the $\MOMi$ and
$\mMOM$ schemes decrease from two loops to three loops and then increase at the
next order, indicating that the convergence can qualitatively differ between 
schemes. One may suggest that because the $\MSbar$ scheme has the smallest 
leading order coupling constant then we expect this scheme to converge quickest
towards the true value of the series. While the loop order change in the 
$\MSbar$ scheme is often smaller than the others, we see in the case of Table 1
that in going from $p$~$=$~$3$ to $p$~$=$~$4$ the data for the $\mMOM$ scheme 
changes by a smaller amount. Equally in Table 2 the $\MOMg$ scheme results 
changed by a smaller amount between the same orders.

We may ask how accurate our averages are at guessing the as yet unknown next 
order. In Table 1 the average values converge with the majority oscillating 
towards the $p$~$=$~$4$ value. It can be seen that at $p$~$=$~$2$ the apparent 
scheme independence means that the two loop average gives a very small error of
$0.005$ on $0.0636$. However the four loop value is $0.0004$ on $0.0605$. So 
the average value quoted at two loops is many standard deviations from the 
central value at four loops. On the other hand the value for 
$p$~$=$~$q$~$=$~$3$ is $0.0594\pm 0.0014$ meaning the $p$~$=$~$q$~$=$~$4$ value
easily lies within the range. 

Next we can examine how much the partial higher order information provided by 
$p$~$<$~$q$ can give about the convergence of the values. There are cases where
the values appear to converge toward the higher loop values, however this is 
not true in general. Expanding $a^{\cal S}_{NN}$ in terms of the leading order 
coupling constant and treating the $\Lambda$-ratio scale change in the 
perturbative scale equation, one finds that the series is identical in all 
schemes to order $a_1^{N}(x)$. Including only an increase in the loop order of 
the coupling constant without additional terms in the series will provide a 
partial contribution to higher orders but there will still be missing terms of 
the order $a_1^{N+1}(x)$. In this expansion the two lowest order terms that 
first appear at order $a^{\cal S}_{NN}$ is the term 
$a_{RR,1}^{\cal S}(s) k_N(Q/\Lambda)a_1^N(Q,\Lambda)$~$=$~$a_{RR,1}^{\cal S}(s)(a^{\cal S}_N -a^{\cal S}_{N-1})$ 
where $k_N(Q/\Lambda)$ is a polynomial in $\ln L^{\cal S}$ and the term  
$a_{RR,N}^{\cal S} a_1^N(Q,\Lambda)$. We can get the first of these without 
calculating the $N$-th term in the R ratio series, so only if this term
dominates, which equates to the condition 
$| a_{RR,1}^{\cal S}(s)k_N(Q/\Lambda)|$~$>$~$|a_{RR,N}^{\cal S}|$, does the
higher order of the coupling constant provide strong information on the 
convergence of the series. Analysis of the table indicates the new term from 
the series either dominates or is commensurate with the new term from the 
coupling constant. 

\subsection{Effective coupling analysis.}

Moving forward in our analysis we will now consider the effective coupling 
constants associated with the R ratio and Bjorken sum rule. Indeed the 
formalism will equally apply to other observables and we assume the 
perturbative series for any of these will be of the form
\begin{eqnarray}
\rho(Q^2) &=& \rho_0 ~+~ \rho_1 a_{\cal S}(Q^2) ~+~ 
\rho_2 a_{\cal S}^2(Q^2) ~+~
\rho_3 a_{\cal S}^3(Q^2) ~+~ \rho_4 a_{\cal S}^4(Q^2) ~+~ \ldots
\end{eqnarray}
where $\rho(Q^2)$ represents the observable. A quantity is still measureable if 
it can be related to another measureable by adding to or multiplying by a 
constant. Therefore we can construct an observable quantity that acts like a 
coupling constant to leading order using
\begin{eqnarray}
a^{\rho}(Q^2) &=&  \frac{\rho(Q^2)-\rho_0}{\rho_1} ~=~ a_{\cal S}(Q^2)
+c_{1}^{\rho,{\cal S}} a_{\cal S}^2(Q^2)
+c_{2}^{\rho,{\cal S}} a_{\cal S}^3(Q^2)
+c_{3}^{\rho,{\cal S}} a_{\cal S}^4(Q^2) + \ldots
\label{eq:effectivecoupling}
\end{eqnarray}
Examining the running of this effective coupling constant across a range of 
energy values within the perturbative regime will allow us to test whether the 
trends found in the Tables are specific to the points considered or are more 
general trends for the series in the range. In order to test our ideas of 
residual scheme dependence in a measureable due to truncation of a perturbative 
series we can plot their running across a range of values. In \cite{29} the 
momentum range chosen was $20 \Lambda^{\MSbars}_{\Nf}$ to 
$200 \Lambda^{\MSbars}_{\Nf}$ which represented a large part of the 
perturbative regime especially when $\Lambda^{\MSbars}_{\Nf}$ was converted to 
more conventional energy units. In general the value of
$\Lambda^{\MSbars}_{\Nf}$ is larger for higher values of $\Nf$. Therefore the 
momentum considered in \cite{29} for more active quarks will be lower. This 
will mean that the effect of truncation will be greater for $\Nf$~$=$~$6$ than 
for $\Nf$~$=$~$3$. This is counter to what is expected phenomenologically. 
However we continue with this range to allow for direct comparison with the 
previous results. It is important to note for instance that due to the 
particular polynomial dependence on $\Nf$ in different scheme perturbative 
series for an observable, when it is numerically evaluated certain terms may 
dominate for one specific number of active quarks but not for another number. 
How this transpires can be seen when, for example, (\ref{rratnum}) is evaluated
for a range of $\Nf$. 

Scheme comparison plots of the R ratio at both $O(a^3)$ and $O(a^4)$ are 
provided in Figures \ref{fig:RR3LScheme} and \ref{fig:RR4LScheme} respectively.
For $\Nf$~$=$~$6$ in the $O(a^4)$ plot we see a very small spread in the value 
of the series across all momenta, but especially at high momenta where the 
coupling constant tends to zero meaning higher order corrections are 
indecipherable. Across the range $20 \Lambda^{\MSbars}$ to 
$200 \Lambda^{\MSbars}$ the $\MOMq$ scheme provides an upper bound on the 
series, but for the upper bound at $20\Lambda^{\MSbars}$ the minimum line is 
that of the $\mMOM$ scheme whereas at $200\Lambda^{\MSbars}$ it is that of the 
$\MOMg$ scheme. Therefore the schemes are differently ordered at different 
energies. As expected at $\Nf$~$=$~$3$ there is a much smaller, but not 
non-existent, spread for all the schemes. This is mirrored by the scale of the 
series in both cases. Note that at no point along the series does $\MOMg$ 
provide the lower bound. So as mentioned the ordering of the series does depend
on the value of $\Nf$. For $O(a^3)$ we see the values of the series are similar
when compared with $O(a^4)$, but provide a much larger spread of values at all 
energy scales. This is especially prevalent for $\Nf$~$=$~$6$, where the 
$\MSbar$ and $\mMOM$ schemes provide bounds, meaning the schemes are not in a 
fixed order between loop orders. 

To ensure that any results derived from these graphs are not dependent on the 
specific observable we have repeated the exercise for the Bjorken sum rule with
the behaviour of the effective coupling presented in Figures 
\ref{fig:BJ3LScheme} and \ref{fig:BJ4LScheme} for $O(a^3)$ and $O(a^4)$
respectively. First considering the $O(a^4)$ graph we see that the schemes 
providing the maximum and minimum lines are not the same with $\MSbar$ being 
the minimum for $\Nf$~$=$~$3$. While the scheme difference is in general lower 
at $O(a^4)$ than $O(a^3)$ we see that the jump is not quite as severe. At 
$O(a^3)$ it is worth noting that many of the series are highly correlated with 
just the $\MOMq$ scheme behaviour being an outlier at $\Nf$~$=$~$6$ and both 
the $\MSbar$ and $\MOMq$ schemes being outliers in the $\Nf$~$=$~$3$ graph.

\begin{figure}[ht]
\includegraphics[width=7.6cm,height=8cm]{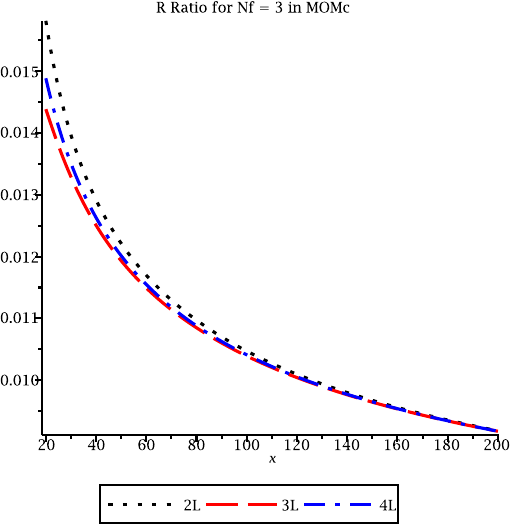}
\quad
\includegraphics[width=7.6cm,height=8cm]{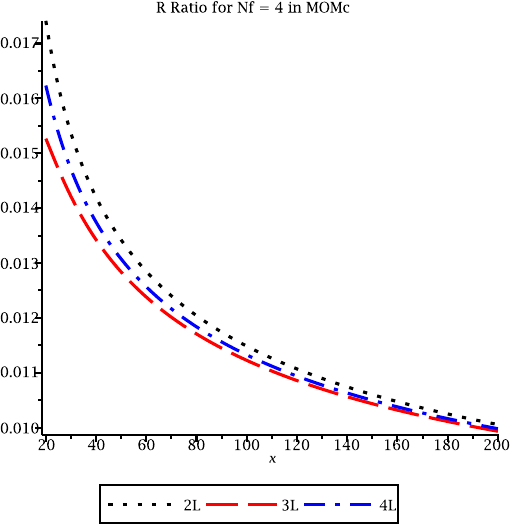}

\vspace{1.2cm}
\includegraphics[width=7.6cm,height=8cm]{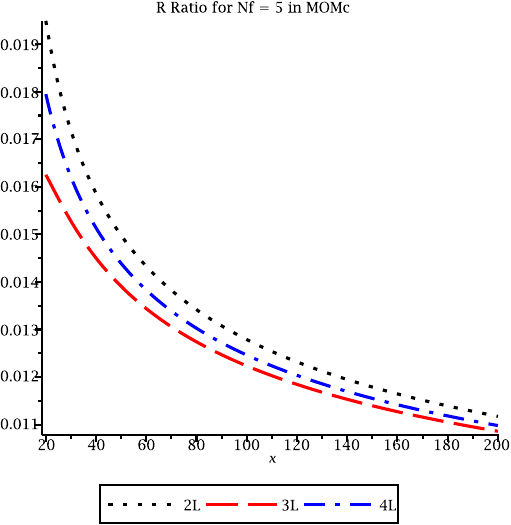}
\quad
\includegraphics[width=7.6cm,height=8cm]{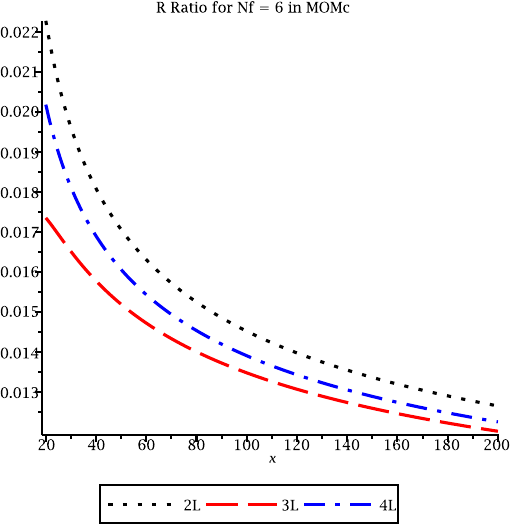}
\caption{Comparison of $a_{RR}^{\MOMcs}(x)$ at various
loops for $\Nf$~$=$~$3$, 
$4$, $5$ and $6$.}\label{fig:RRMOMcLoop}
\end{figure}

\begin{figure}[ht]
\includegraphics[width=7.6cm,height=8cm]{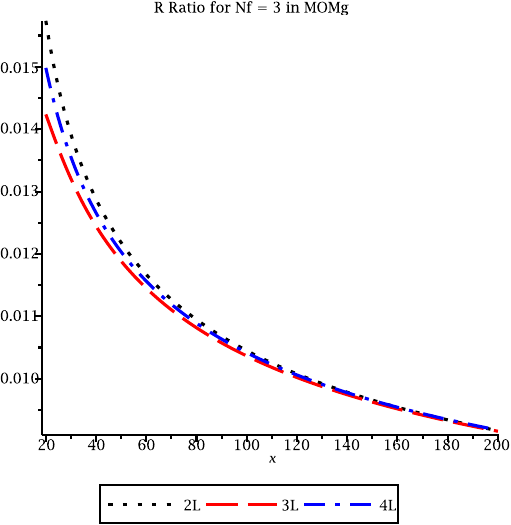}
\quad
\includegraphics[width=7.6cm,height=8cm]{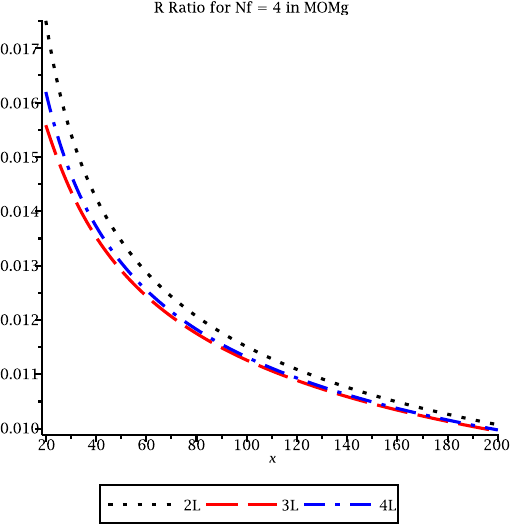}

\vspace{1.2cm}
\includegraphics[width=7.6cm,height=8cm]{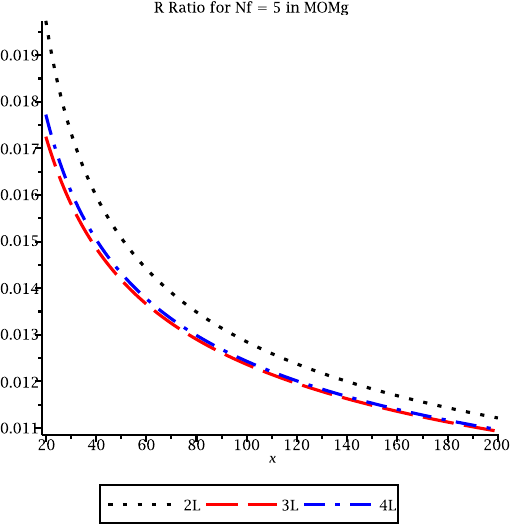}
\quad
\includegraphics[width=7.6cm,height=8cm]{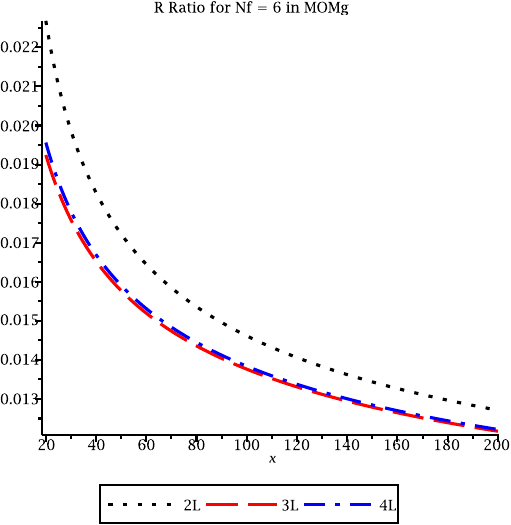}
\caption{Comparison of $a_{RR}^{\MOMgs}(x)$ at various loops for 
$\Nf$~$=$~$3$, $4$, $5$ and $6$.}
\label{fig:RRMOMgLoop}
\end{figure}

\begin{figure}[ht]
\includegraphics[width=7.6cm,height=8cm]{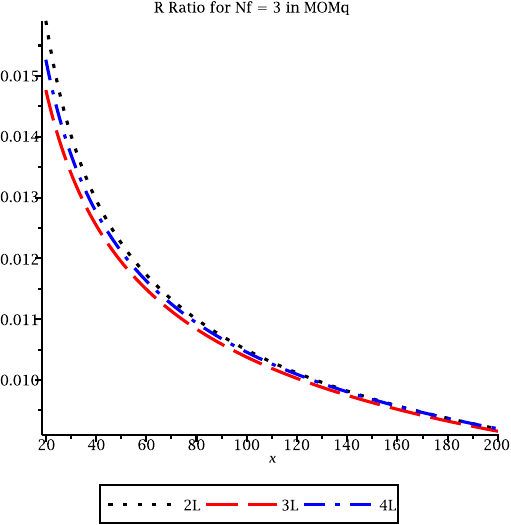}
\quad
\includegraphics[width=7.6cm,height=8cm]{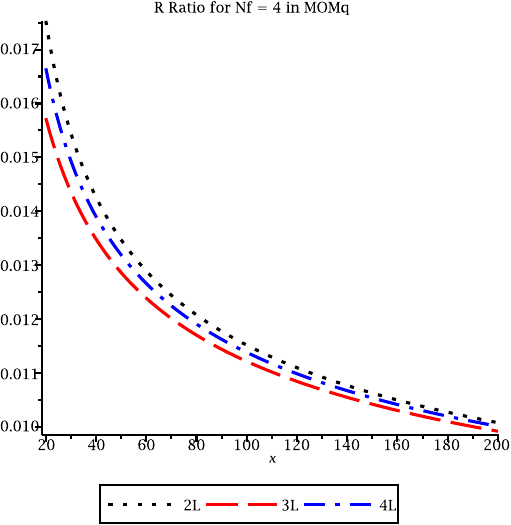}

\vspace{1.2cm}
\includegraphics[width=7.6cm,height=8cm]{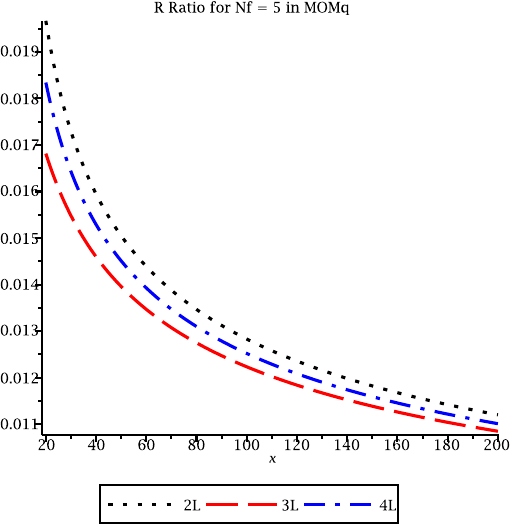}
\quad
\includegraphics[width=7.6cm,height=8cm]{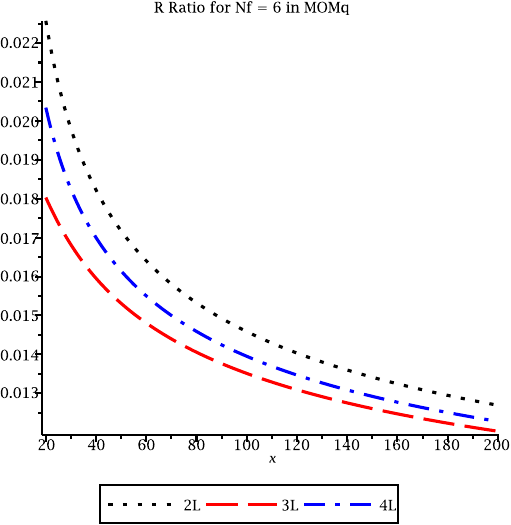}
\caption{Comparison of $a_{RR}^{\MOMqs}(x)$ at various
loops for $\Nf$~$=$~$3$, $4$, $5$ and $6$.}
\label{fig:RRMOMqLoop}
\end{figure}

\begin{figure}[ht]
\includegraphics[width=7.6cm,height=8cm]{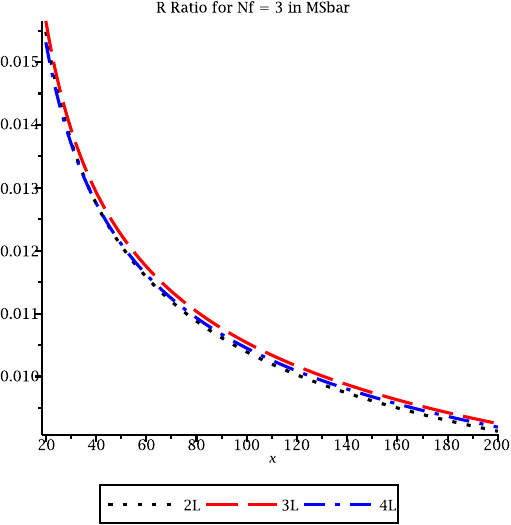}
\quad
\includegraphics[width=7.6cm,height=8cm]{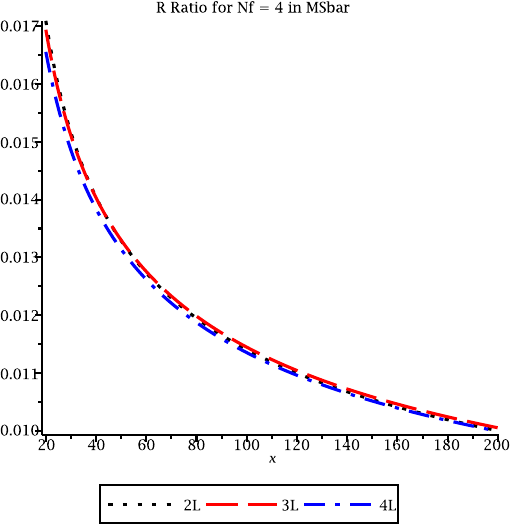}

\vspace{1.2cm}
\includegraphics[width=7.6cm,height=8cm]{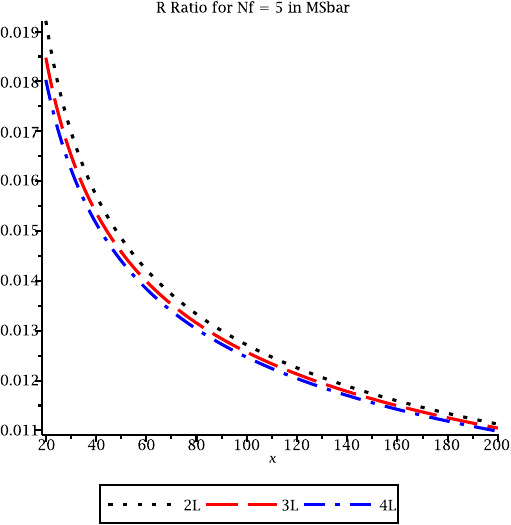}
\quad
\includegraphics[width=7.6cm,height=8cm]{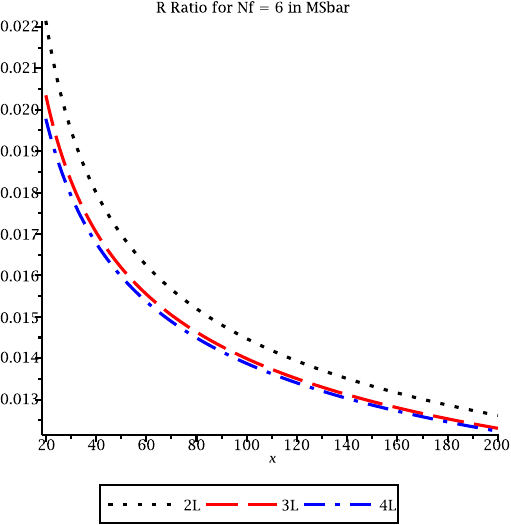}
\caption{Comparison of $a_{RR}^{\MSbars}(x)$ at various
loops for $\Nf$~$=$~$3$, $4$, $5$ and $6$.}
\label{fig:RRMSbarLoop}
\end{figure}

\begin{figure}[ht]
\includegraphics[width=7.6cm,height=8cm]{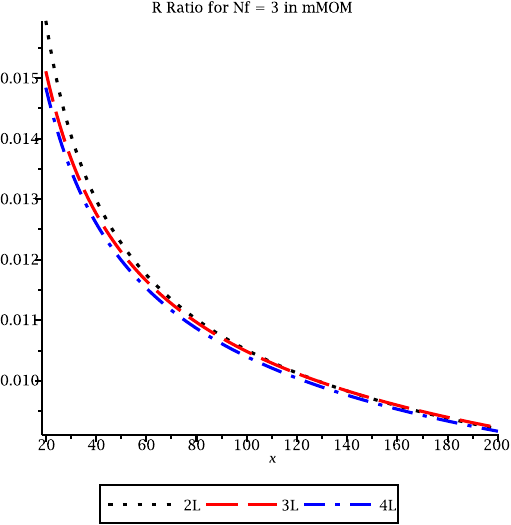}
\quad
\includegraphics[width=7.6cm,height=8cm]{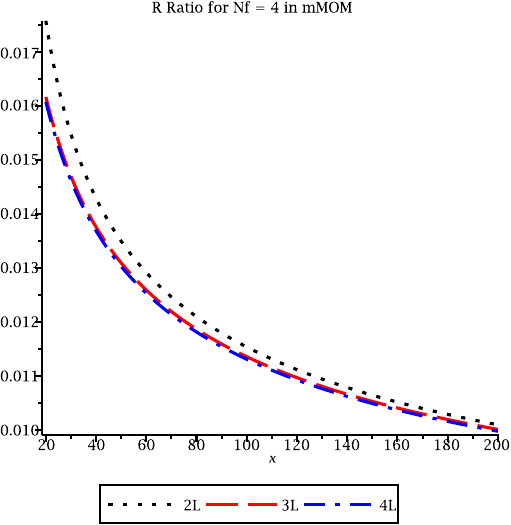}

\vspace{1.2cm}
\includegraphics[width=7.6cm,height=8cm]{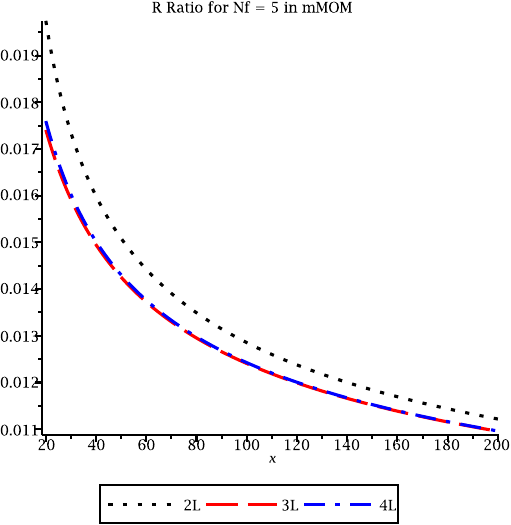}
\quad
\includegraphics[width=7.6cm,height=8cm]{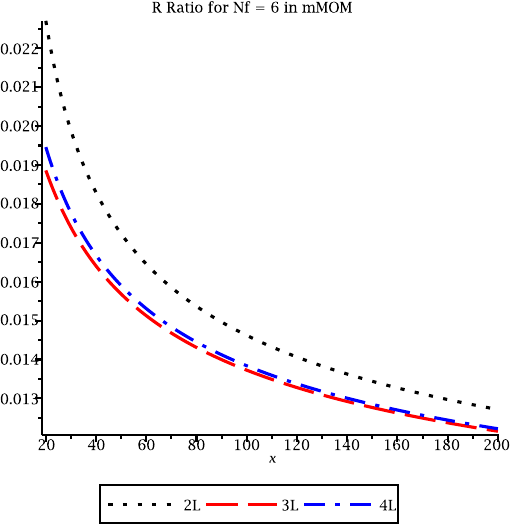}
\caption{Comparison of $a_{RR}^{\mMOM}(x)$ at various
loops for $\Nf$~$=$~$3$, $4$, $5$ and $6$.}
\label{fig:RRmMOMLoop}
\end{figure}

\begin{figure}[ht]
\includegraphics[width=7.6cm,height=8cm]{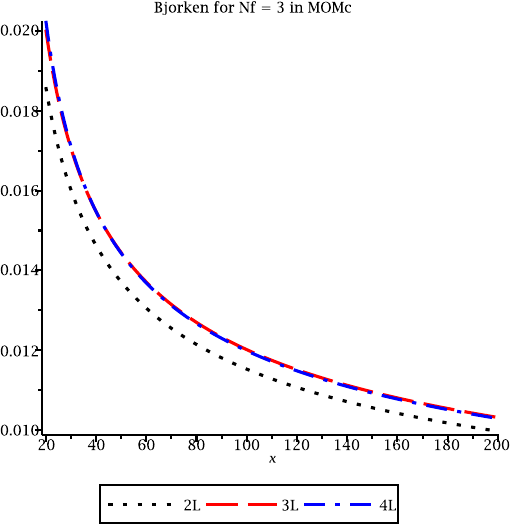}
\quad
\includegraphics[width=7.6cm,height=8cm]{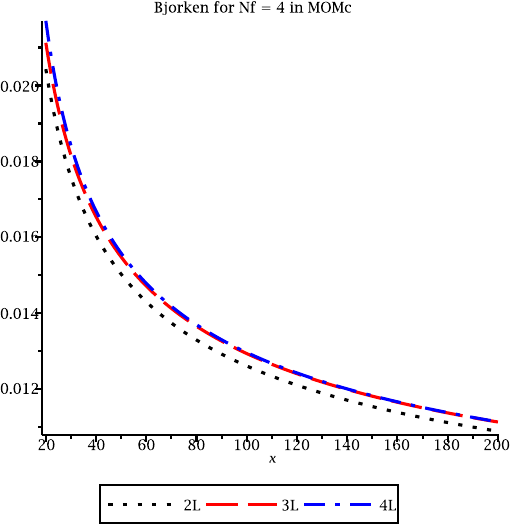}

\vspace{1.2cm}
\includegraphics[width=7.6cm,height=8cm]{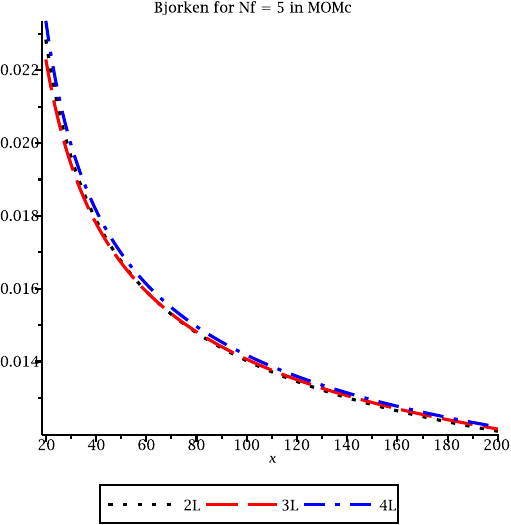}
\quad
\includegraphics[width=7.6cm,height=8cm]{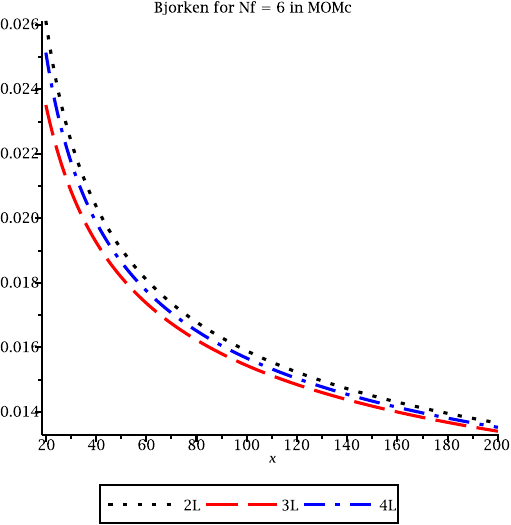}
\caption{Comparison of $a_{{\mbox{\footnotesize{Bjr}}}}^{\MOMcs}(x)$ at 
various loops for $\Nf$~$=$~$3$, $4$, $5$ and $6$.} \label{fig:BJMOMcLoop}
\end{figure}

\begin{figure}[ht]
\includegraphics[width=7.6cm,height=8cm]{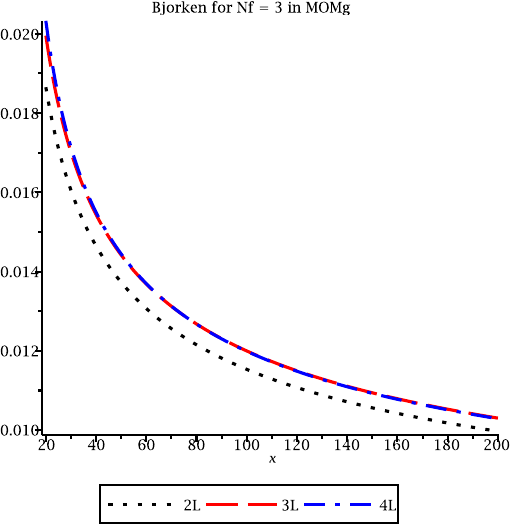}
\quad
\includegraphics[width=7.6cm,height=8cm]{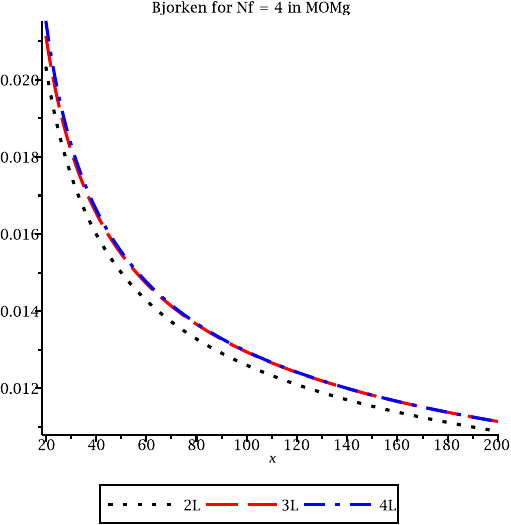}

\vspace{1.2cm}
\includegraphics[width=7.6cm,height=8cm]{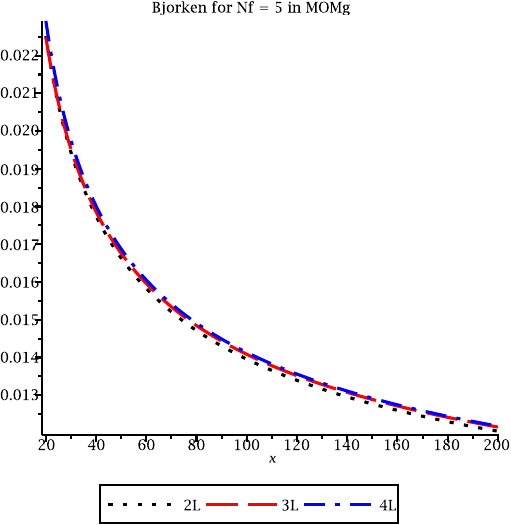}
\quad
\includegraphics[width=7.6cm,height=8cm]{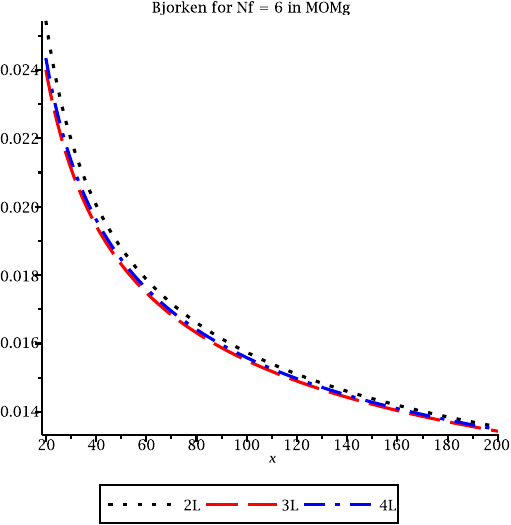}
\caption{Comparison of $a_{{\mbox{\footnotesize{Bjr}}}}^{\MOMgs}(x)$ at 
various loops for $\Nf$~$=$~$3$, $4$, $5$ and $6$.}\label{fig:BJMOMgLoop}
\end{figure}

\begin{figure}[ht]
\includegraphics[width=7.6cm,height=8cm]{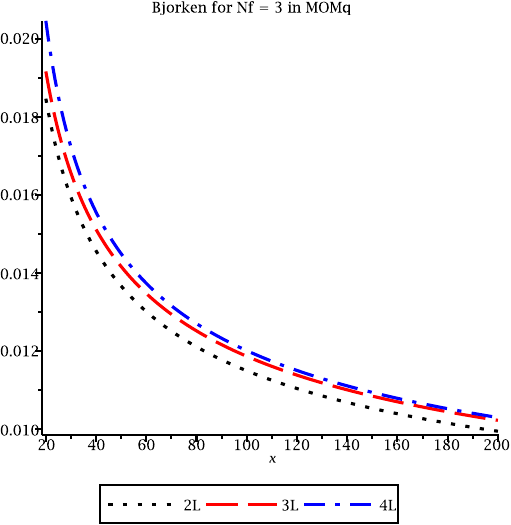}
\quad
\includegraphics[width=7.6cm,height=8cm]{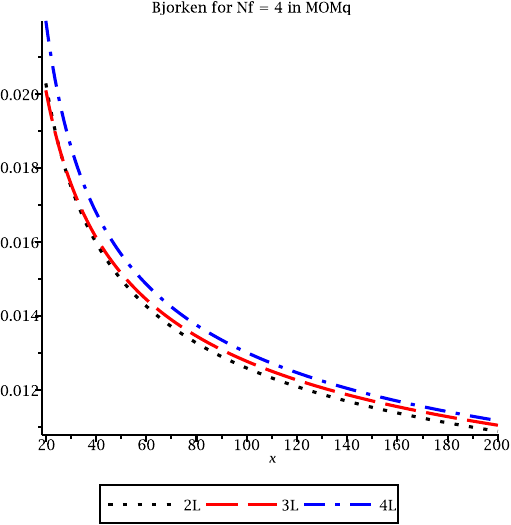}

\vspace{1.2cm}
\includegraphics[width=7.6cm,height=8cm]{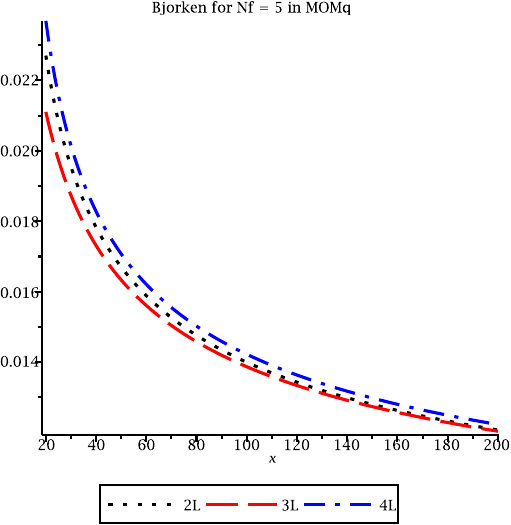}
\quad
\includegraphics[width=7.6cm,height=8cm]{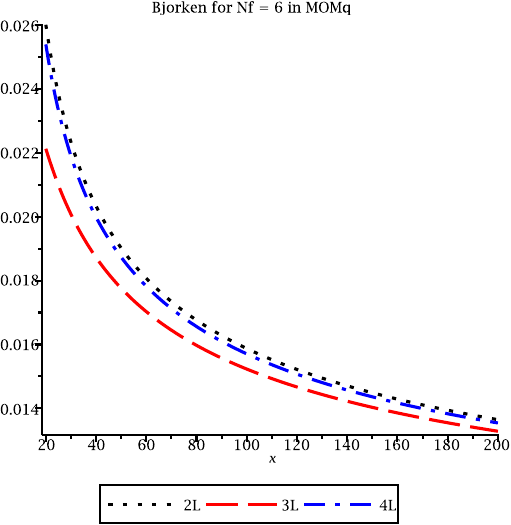}
\caption{Comparison of $a_{{\mbox{\footnotesize{Bjr}}}}^{\MOMqs}(x)$ at 
various loops for $\Nf$~$=$~$3$, $4$, $5$ and $6$.}
\label{fig:BJMOMqLoop}
\end{figure}

\begin{figure}[ht]
\includegraphics[width=7.6cm,height=8cm]{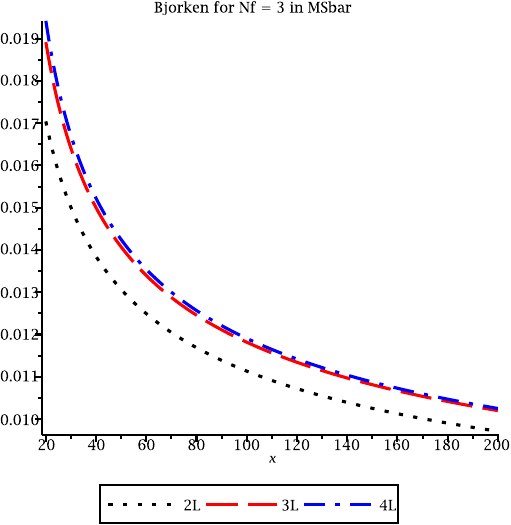}
\quad
\includegraphics[width=7.6cm,height=8cm]{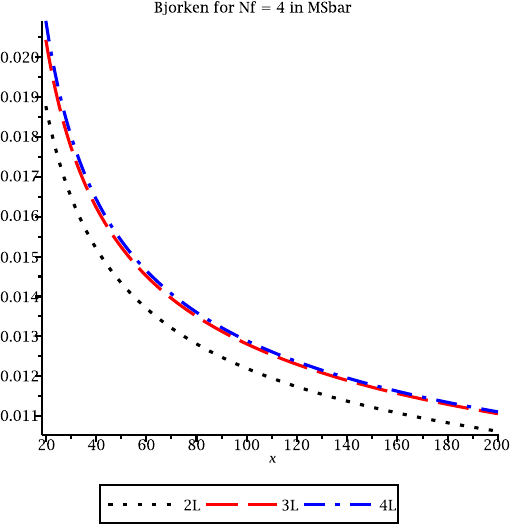}

\vspace{1.2cm}
\includegraphics[width=7.6cm,height=8cm]{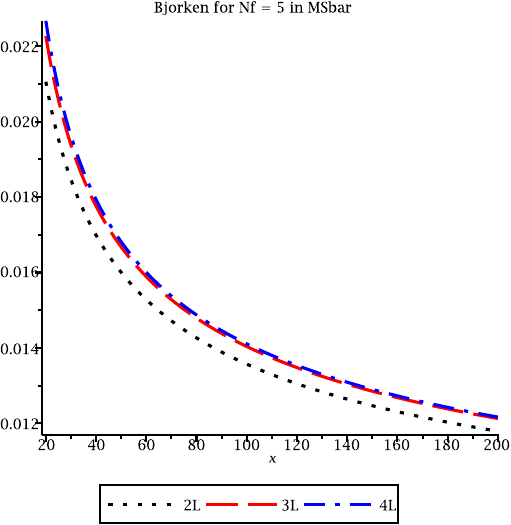}
\quad
\includegraphics[width=7.6cm,height=8cm]{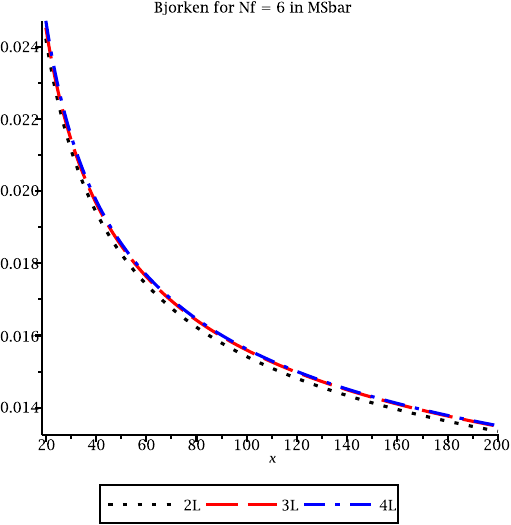}
\caption{Comparison of $a_{{\mbox{\footnotesize{Bjr}}}}^{\MSbars}(x)$ at
various loops for $\Nf$~$=$~$3$, $4$, $5$ and $6$.}
\label{fig:BJMSbarLoop}
\end{figure}

\begin{figure}[ht]
\includegraphics[width=7.6cm,height=8cm]{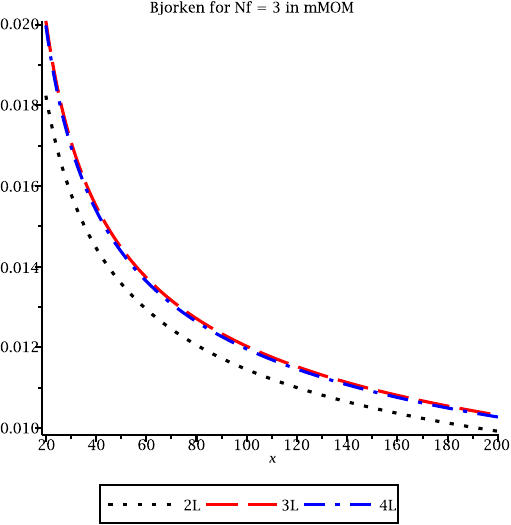}
\quad
\includegraphics[width=7.6cm,height=8cm]{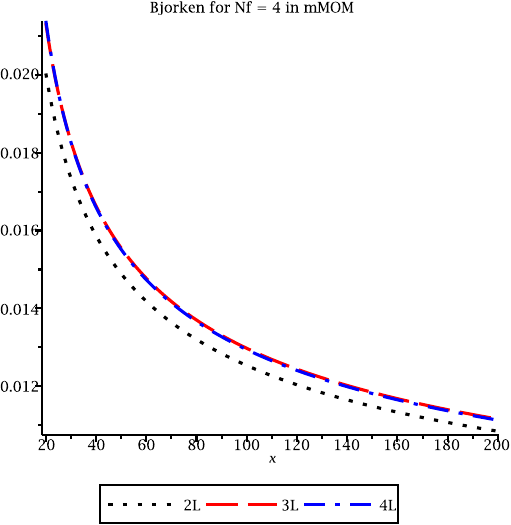}

\vspace{1.2cm}
\includegraphics[width=7.6cm,height=8cm]{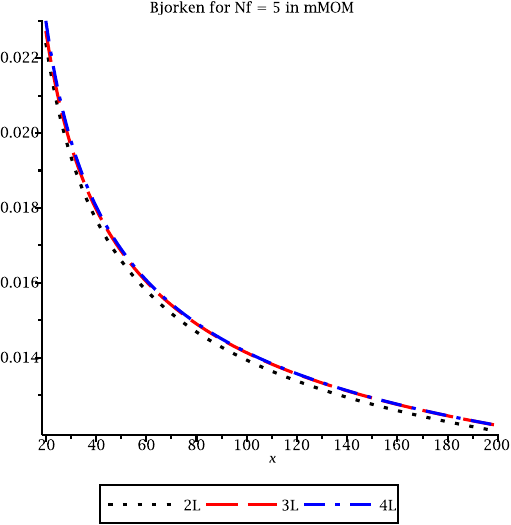}
\quad
\includegraphics[width=7.6cm,height=8cm]{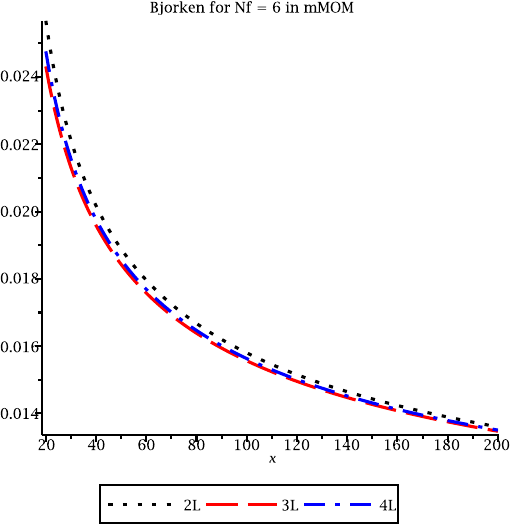}
\caption{Comparison of $a_{{\mbox{\footnotesize{Bjr}}}}^{\mMOM}(x)$ at various
loops for $\Nf$~$=$~$3$, $4$, $5$ and $6$.}\label{fig:BJmMOMLoop}
\end{figure}

An alternative way of quantifying the behaviour of the observable series is to
consider the development of the effective coupling with loop order.
Consequentially this will allow for a deeper understanding of how each series 
converges towards the natural value for each scheme. By plotting the series 
calculated in a single scheme but at different loop orders together we can gain
an understanding of the more general qualitative features of the convergence of
each scheme across a larger range. Therefore in Figures \ref{fig:RRMOMcLoop}, 
\ref{fig:RRMOMgLoop}, \ref{fig:RRMOMqLoop}, \ref{fig:RRMSbarLoop} and 
\ref{fig:RRmMOMLoop} we have provided loop comparisons of the R ratio in the 
$\MOMc$, $\MOMg$, $\MOMq$, $\MSbar$ and $\mMOM$ schemes, respectively, for four
values of $\Nf$. In the case of $\MOMc$ scheme we see the series oscillates as 
the loop order increases since the $O(a^4)$ line sits between the two lower 
order series. The spread of this oscillation decreases with $\Nf$. This 
behaviour is shared by all other $\MOMi$ schemes, whereas the $\MSbar$ and 
$\mMOM$ schemes decrease alternating at each order but converging towards a 
value. In addition the same loop comparisons in each scheme are given for the 
Bjorken sum rule in Figures \ref{fig:BJMOMcLoop}, \ref{fig:BJMOMgLoop}, 
\ref{fig:BJMOMqLoop}, \ref{fig:BJMSbarLoop} and \ref{fig:BJmMOMLoop}. Focussing
first on the $\MOMc$ convergence, for $\Nf$~$=$~$6$ we see the series 
oscillates. While this is also seen at high momentum for $\Nf$~$=$~$5$, at low 
momenta this changes at $O(a^3)$ in the middle which suggests the breakdown of
perturbation theory. Finally for $\Nf$~$=$~$3$ we see that the $O(a^3)$ and 
$O(a^4)$ cases are nearly identical but the $O(a^2)$ line lies much further 
below. This general behaviour is shared by the other $\MOMi$ schemes. 

\sect{Scheme error.}\label{sec:Scheme Error}

In quantum field theory typically the calculation of a measureable quantity has 
no closed form solution. So one must use an approximation often given by a
perturbative series in some small quantity. In order to quantify the accuracy 
of these approximations it is important to assign an error measurement to them. 
In perturbative QCD measureable quantities are expected to be renormalization 
group invariant because results in Nature should not be dependent on how we 
choose to calculate the quantity. One way this is embodied is in scale 
invariance in which the calculation of a physical observable is expected to be 
independent of the choice of the renormalization scale. This idea is used to 
calculate an error estimate with conventional scale setting commonly employed
which involves varying the scale, typically by halving and doubling some
representative scale of the process. The magnitude the scale is varied by is 
arbitrary but dictated by what has worked in the past as opposed to some deeper
theoretical understanding. It is therefore worthwhile comparing this method to 
other estimates of the accuracy of a series. Recent reviews of methods to fix 
scheme and scale ambiguity are given in \cite{49} and \cite{73}. However here 
we will be using this ambiguity to quantify the uncertainty in theoretical 
calculations. Renormalization group invariance also implies the invariance of 
measureable quantities with respect to the choice of renormalization scheme when
evaluated to all orders. In this section we formalise the ideas discussed 
previously to investigate, for contrast, the use of scheme difference as a 
measure of theoretical error in perturbative QCD. Examples of the use of scheme
error in calculations include \cite{51} and \cite{52}. For this study we will 
assume we have a well-defined renormalization scale. 

First we suppose that a measureable quantity $\rho(Q)$ at scale $Q$ has been
evaluated up to the $N$-loop level
\begin{eqnarray}
\rho(Q) &=& \sum_{j=0}^N \rho^{\cal S}_ja_{\cal S}(Q)^j ~+~ 
\Delta(\rho^{\cal S},N,Q) ~=~ \rho^{\cal S}_{(N)}(Q) ~+~
\Delta(\rho^{\cal S},N,Q)
\label{eq:SeriesError}
\end{eqnarray}
where $\Delta(\rho^{\cal S},N,Q)$ represents the difference between the finite 
approximation and the true value of the series. Perturbatively it will be 
$O(a^{N+1})$. Calculating this in schemes ${\cal S}_1$ and ${\cal S}_2$ gives
\begin{eqnarray}
\rho(Q) &=& \rho^{{\cal S}_1}_{(N)}(Q) ~+~ \Delta(\rho^{{\cal S}_1},N,Q) ~=~
\rho^{{\cal S}_2}_{(N)}(Q) ~+~ \Delta(\rho^{{\cal S}_2},N,Q) ~.
\end{eqnarray}
This has two distinct cases, either:
\begin{itemize}
\item 
$\mathbf{sgn}(\Delta(\rho^{{\cal S}_1},N,Q))$~$=$~$\mathbf{sgn}
(\Delta(\rho^{{\cal S}_2},N,Q))$ in which case both 
$\rho^{{\cal S}_1}_{(N)}(Q)$ and $\rho^{{\cal S}_2}_{(N)}(Q)$ will produce 
over- or under-estimates of the true value of the series,
\item 
$\mathbf{sgn}(\Delta(\rho^{{\cal S}_1},N,Q))$~$\neq$~$\mathbf{sgn}
(\Delta(\rho^{{\cal S}_2},N,Q))$ in which case $\rho^{{\cal S}_1}_{(N)}(Q)$ and
$\rho^{{\cal S}_2}_{(N)}(Q)$ will act as bounds on the true value of $\rho(Q)$.
\end{itemize} 

Since in general $\Delta(\rho^{\cal S},N,Q)$ is not known, as this would be 
equivalent to knowing the true value of the series, we cannot know which case a
particular pair of schemes will fit into. However, if we consider more schemes 
there will be a greater chance that the scheme with the maximum value and the 
scheme with the minimum value will correctly bound the true value of the 
series, provided there is no reason to expect additional correlation between 
the schemes as in the case at the two loop level.

\subsection{Scheme envelopes.}

Thus far we have discussed how our measure of error may bound the true value 
without assigning a degree of belief to that bound, a concept which is 
discussed at length in \cite{74}. One could attempt to consider the values of 
the schemes to exist in some probability distribution which could then be used 
to calculate a mean and standard deviation on the series. However no 
distribution is known. In Section \ref{sec:Analysis} we used the arithmetic 
mean and unbiased standard deviation which allowed us to provide a rough 
estimate of the true value of the series and remaining error suggested by the 
values calculated in each scheme. This had the advantage of allowing one to use
the typical interpretation of the standard deviation as a degree of belief in 
the value reported. However, the use of the arithmetic mean required the 
assumption that all schemes are independent which due to correlation of the 
scheme definitions, particularly for the $\MOMi$ schemes, is not an accurate 
assumption. This can be further demonstrated by the qualitative similarity of 
the convergence of the $\MOMi$ schemes and illustrated in the loop comparison 
graphs. In this subsection we aim to discuss the idea of scheme error more 
concretely and therefore we have chosen not to make this assumption. So we will
not attach any degree of belief to our error. 

   \begin{table}[ht]
\begin{center}
 \begin{tabular}{ |c|c|c| }
 \hline 
 \multicolumn{3}{|c|}{$\alpha_s^{RR}=0.16556 \pm 0.01571$}\\ 
 \multicolumn{3}{|c|}{at $Q= 31.62278$ GeV} \\ 
 \hline 
 \hline 
 &&\\ Scheme & $\mbox{\L}$  &$\alpha_s^{\MSbars}(M_Z)$ \\ & & \\ 
 \hline  \rule{0pt}{12pt} 
  & 2 & $ 0.12772^{ +0.00948 }_{-0.00904}$  \\ 
  $\MSbar$ & 3 & $ 0.13084^{ +0.01019 }_{-0.00986}$  \\ 
  & 4 & $ 0.13187^{ +0.01050 }_{-0.01027}$ \\ 
 \hline 
  &2 & $ 0.12662^{ +0.00923 }_{-0.00875}$  \\ 
 $\mMOM$  & 3 & $ 0.13260^{ +0.01083 }_{-0.01081}$  \\ 
  & 4 & $ 0.13238^{ +0.01072 }_{-0.01063}$  \\ 
 \hline 
  & 2 & $ 0.12680^{ +0.00926 }_{-0.00879}$  \\ 
 $\MOMq$ & 3 & $ 0.13460^{ +0.01148 }_{-0.01171}$  \\ 
  & 4 & $ 0.13131^{ +0.01030 }_{-0.00999}$ \\ 
 \hline 
   & 2 & $ 0.12665^{ +0.00923 }_{-0.00875}$  \\ 
 $\MOMg$ & 3 & $ 0.13306^{ +0.01097 }_{-0.01102}$  \\ 
  & 4 & $ 0.13228^{ +0.01067 }_{-0.01054}$  \\ 
 \hline 
  & 2& $ 0.12713^{ +0.00934 }_{-0.00888}$  \\ 
 $\MOMc$  & 3 & $ 0.13479^{ +0.01171 }_{-0.01239}$  \\ 
  & 4& $ 0.13194^{ +0.01053 }_{-0.01032}$  \\ 
 \hline \hline 
 & 2 & $0.12717 \pm 0.00055^{ +0.00948 }_{-0.00875}$\\ 
 Average & 3 & $0.13281 \pm 0.00197^{ +0.01171 }_{-0.00986}$ \\ 
 & 4& $0.13185 \pm 0.00053^{ +0.01072 }_{-0.00999}$\\ \hline 
 \end{tabular} 
 \end{center}\caption{{Estimates of $\alpha_s^{\MSbars} (M_Z)$ from the 
 R ratio effective coupling at order $\mbox{\L}$ using data from \cite{75} 
 calculated 
 in the $\MSbar$, $\MOMi$ and $\mMOM$ schemes. The error on the average is the 
 envelope of the scheme values and the average value is the centre of this
 envelope.}
 \label{tab:RRDATA1}}
 \end{table}

To give a concrete example we recall that the R ratio effective coupling 
constants with five active fermions in our five schemes are given numerically
by 
\begin{eqnarray}
a_{RR}^{\mMOMs}\vert_{N_f=5}&=&a_{\mMOMs} ~-~ 2.89086a_{\mMOMs}^2 ~-~ 
299.03413a_{\mMOMs}^3 ~+~ 407.37433a_{\mMOMs}^4 \nonumber \\
&& +~ O(a_{\mMOMs}^5) \nonumber \\
a_{RR}^{\MOMcs}\vert_{N_f=5}&=&a_{\MOMcs} ~-~ 7.35580a_{\MOMcs}^2 ~-~ 
258.15390a_{\MOMcs}^3 ~+~ 3883.57250a_{\MOMcs}^4 \nonumber \\
&& +~ O(a_{\MOMcs}^5) \nonumber \\
a_{RR}^{\MOMgs}\vert_{N_f=5}&=&a_{\MOMgs} ~-~ 3.77154a_{\MOMgs}^2 ~-~ 
275.52688a_{\MOMgs}^3 ~+~ 1671.72909a_{\MOMgs}^4 \nonumber \\
&& +~ O(a_{\MOMgs}^5) \nonumber \\
a_{RR}^{\MOMqs}\vert_{N_f=5}&=&a_{\MOMqs} ~-~ 5.52330a_{\MOMqs}^2 ~-~ 
168.45783a_{\MOMqs}^3 ~+~ 1337.29074a_{\MOMqs}^4 \nonumber \\
&& +~ O(a_{\MOMqs}^5) \nonumber \\
a_{RR}^{\MSbars}\vert_{N_f=5}&=&a_{\MSbars} ~+~ 5.63692a_{\MSbars}^2 ~-~ 
204.27304a_{\MSbars}^3 ~-~ 5118.76040a_{\MSbars}^4 \nonumber \\
&& +~ O(a_{\MSbars}^5) ~.
\label{eq:aRRMSBARNF5}
\end{eqnarray} 
These have been derived from the R ratio perturbative series using the notation
of (\ref{eq:effectivecoupling}). For a valid perturbative series at the $N$-th 
loop level we expect $\Delta(\rho^{\cal S},N,Q)$ to be dominated by the 
$(N+1)$-th term in the original series. By considering the sign of the third 
term in each series we get a rough estimate of the $\Delta$-value at the two 
loop level for instance. Since each sign is negative this means that each value
calculated at $O(a^2)$ will be an over-estimate of the series, which explains 
the lower scheme difference at the two loop level when compared to three loops. 
This should provide an accurate error since the signs are different on the 
$O(a^4)$ terms in each series. Since we do not know the five loop terms in any 
series the signs of the $\Delta$-values at the four loop level cannot be known 
exactly. However there is no reason to expect all five schemes to share the 
same sign at that loop level. We can therefore consider the envelope provided 
by the $O(a^3)$ lines of the plots as an absolute limit on the true value of 
the series with the lines provided by the $O(a^4)$ values as the potential 
minimum error at that order. We note that this argument is simplified for 
clarity as it ignores the modification of the coupling constant at each order. 
Although as discussed previously the dominant term is typically from the new 
term in the series. 

\begin{table}[ht]
\begin{center}
 \begin{tabular}{ |c|c|c| }
 \hline 
 \multicolumn{3}{|c|}{$\alpha_s^{RR}=0.14546 \pm 0.01382$}\\ 
 \multicolumn{3}{|c|}{at $Q= 59.16080$ GeV} \\ 
 \hline 
 \hline 
 &&\\ Scheme & $\mbox{\L}$ &$\alpha_s^{\MSbars}(M_Z)$ \\ & & \\ 
 \hline  \rule{0pt}{12pt} 
  & 2 & $ 0.12731^{ +0.01070 }_{-0.01045}$  \\ 
  $\MSbar$ & 3 & $ 0.13004^{ +0.01136 }_{-0.01122}$  \\ 
  & 4 & $ 0.13085^{ +0.01162 }_{-0.01156}$ \\ 
 \hline 
  & 2 & $ 0.12636^{ +0.01047 }_{-0.01017}$  \\ 
 $\mMOM$  & 3 & $ 0.13124^{ +0.01180 }_{-0.01185}$  \\ 
  & 4 & $ 0.13116^{ +0.01175 }_{-0.01176}$  \\ 
 \hline 
  & 2 & $ 0.12651^{ +0.01051 }_{-0.01021}$  \\ 
 $\MOMq$ & 3 & $ 0.13273^{ +0.01229 }_{-0.01252}$  \\ 
  & 4 & $ 0.13048^{ +0.01148 }_{-0.01136}$ \\ 
 \hline 
   & 2 & $ 0.12638^{ +0.01047 }_{-0.01017}$  \\ 
 $\MOMg$ & 3 & $ 0.13159^{ +0.01191 }_{-0.01200}$  \\ 
  & 4 & $ 0.13111^{ +0.01172 }_{-0.01172}$  \\ 
 \hline 
  & 2 & $ 0.12680^{ +0.01058 }_{-0.01029}$  \\ 
 $\MOMc$  &  3 & $ 0.13264^{ +0.01233 }_{-0.01269}$  \\ 
  & 4 & $ 0.13089^{ +0.01163 }_{-0.01159}$  \\ 
 \hline \hline 
 & 2 & $0.12683 \pm 0.00048^{ +0.01070 }_{-0.01017}$\\ 
 Average & 3 & $0.13138 \pm 0.00135^{ +0.01229 }_{-0.01122}$ \\ 
 & 4& $0.13082 \pm 0.00034^{ +0.01175 }_{-0.01136}$\\ \hline 
 \end{tabular} 
 \end{center}\caption{{Estimates of $\alpha_s^{\MSbars} (M_Z)$ from the 
 R ratio effective coupling at order $\mbox{\L}$ using data from \cite{75} 
 calculated 
 in the $\MSbar$, $\MOMi$ and $\mMOM$ schemes. The error on the average is the 
 envelope of the scheme values and the average value is the centre of this 
 envelope.}
 \label{tab:RRDATA2}}
 \end{table}

\begin{figure}[ht]
\includegraphics[width=7.6cm,height=8cm]{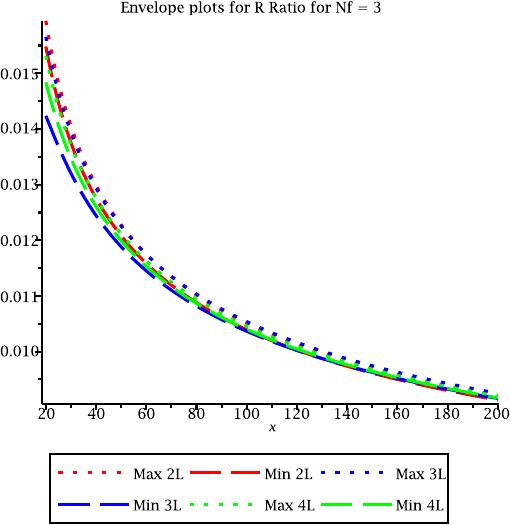}
\quad
\includegraphics[width=7.6cm,height=8cm]{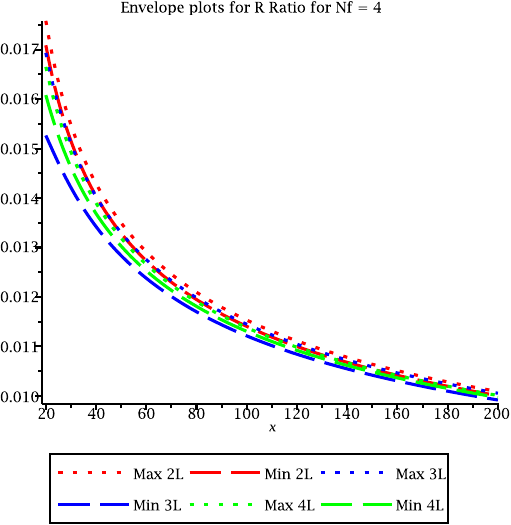}

\vspace{1.2cm}
\includegraphics[width=7.6cm,height=8cm]{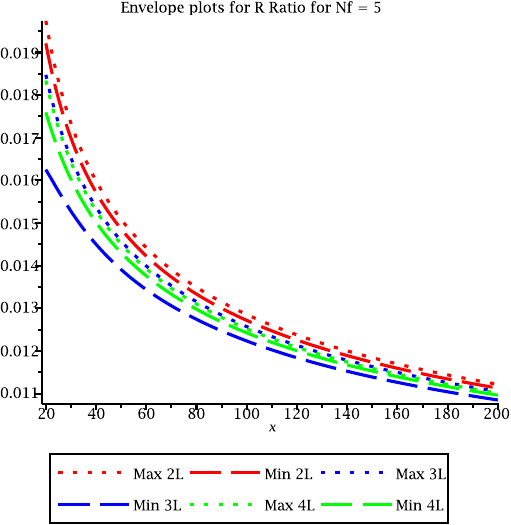}
\quad
\includegraphics[width=7.6cm,height=8cm]{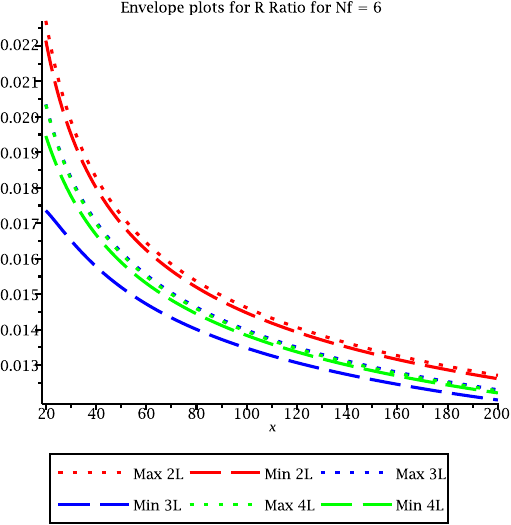}
\caption{Envelope error for $a_{RR}^{\cal S}(x)$ at different 
loop orders for $\Nf$~$=$~$3$, $4$, $5$ and $6$ for the schemes 
considered here.}
\label{fig:EnvelopeRRLoop}
\end{figure}

With this in mind we compare the envelopes for the R ratio effective coupling 
constants at the different loop orders against each other in Figure 
\ref{fig:EnvelopeRRLoop}. At each energy scale all schemes are considered but 
only the ones that evaluate to give the minimum and maximum values are plotted 
at each point. We note that as discussed in Section \ref{sec:Analysis} the 
bounding schemes can change between loop order, at different energies and for 
different $\Nf$. This means that all schemes must be considered at all values 
for an accurate envelope to be formed. For the case of six active fermions we 
see that the two loop envelope sits entirely above the three loop envelope 
which houses the four loop at its upper bound. The fact that there is no 
overlap between the three and four loop cases is explained by the fact that the
three loop term acts to reduce the value of the R ratio in all of the schemes 
considered here. For the situation with three active fermions there exists an 
overlap between the three loop and four loop graphs, although at lower energies
while the four loop lies inside the three loop it is not within the two loop 
envelope. At higher energies however the two and three loop envelopes lie on 
top of each other with the four loop one lying in between. In general we see 
that the two loop and four loop envelopes are of similar size, with that at 
three loops giving a larger bound. 
 
\begin{figure}[hb]
\includegraphics[width=7.6cm,height=8cm]{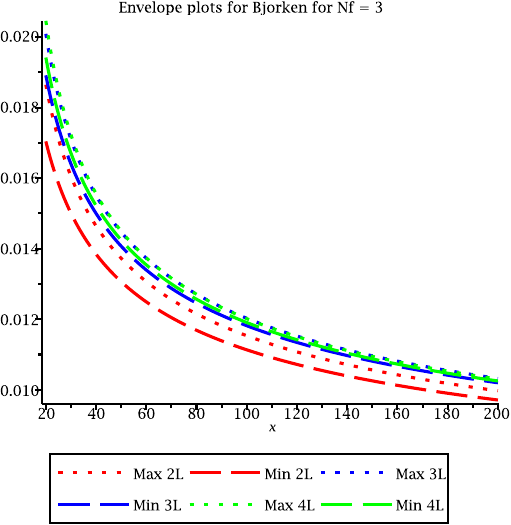}
\quad
\includegraphics[width=7.6cm,height=8cm]{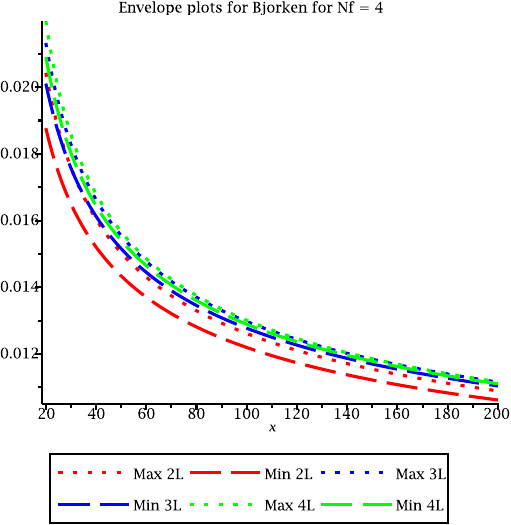}

\vspace{1.2cm}
\includegraphics[width=7.6cm,height=8cm]{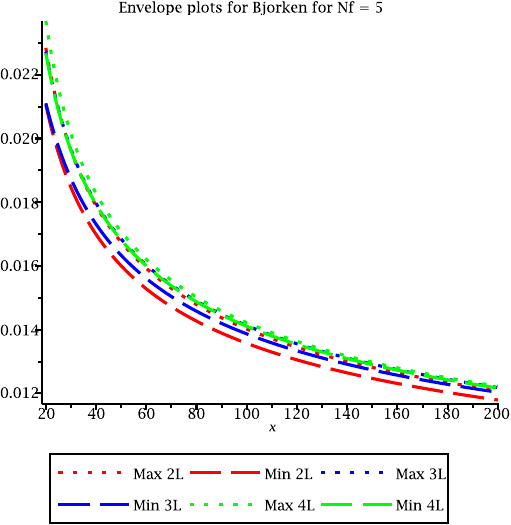}
\quad
\includegraphics[width=7.6cm,height=8cm]{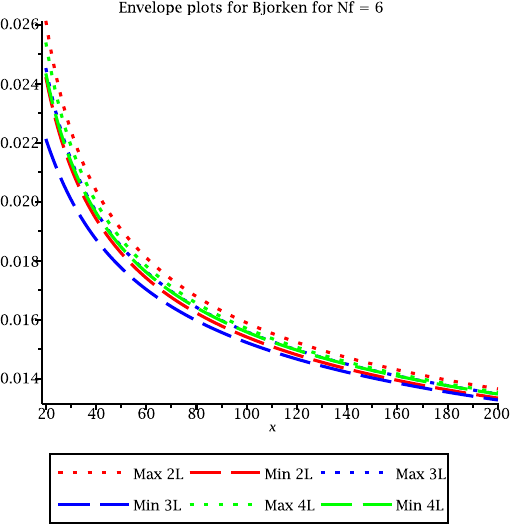}
\caption{Envelope error for $a_{{\mbox{\footnotesize{Bjr}}}}^{\cal S}(x)$ at 
different 
loop orders for $\Nf$~$=$~$3$, $4$, $5$ and $6$ for the schemes 
considered here.}
\label{fig:EnvelopeBJLoop}
\end{figure}

This can be contrasted with similar plots for the Bjorken sum rule given in 
Figure \ref{fig:EnvelopeBJLoop}. Starting with the case of three active 
fermions the $O(a^2)$ error is entirely below and much larger than the 
$O(a^3)$ bound. The full $O(a^4)$ envelope is within the $O(a^3)$ envelope at 
higher energies but at lower energies the upper bound just escapes. As the 
number of active fermions is increased both the two loop and four loop lines
increase relative to the three loop one. So for six active fermions the two 
loop envelope lies almost entirely above the three loop one and the four loop 
lower bound sits on the upper bound of the three loop. In that case the four 
loop one is encapsulated by the two loop one but not the three loop one, 
suggesting that the two loop case gives a more accurate maximum envelope than 
the three loop one.

\begin{table}[ht]
\begin{center}
 \begin{tabular}{ |c|c|c| }
 \hline 
 \multicolumn{3}{|c|}{$\alpha_s^{RR}=0.13697 \pm 0.01225$}\\ 
 \multicolumn{3}{|c|}{at $Q= 82.15838$ GeV} \\ 
 \hline 
 \hline 
 &&\\ Scheme & $\mbox{\L}$ &$\alpha_s^{\MSbars}(M_Z)$ \\ & & \\ 
 \hline  \rule{0pt}{12pt} 
  & 2 & $ 0.12725^{ +0.01066 }_{ -0.01053}$  \\ 
  $\MSbar$ & 3 & $ 0.12982^{ +0.01126 }_{ -0.01124}$  \\ 
  & 4 & $ 0.13056^{ +0.01149 }_{ -0.01154}$ \\ 
 \hline 
  & 2 & $ 0.12635^{ +0.01044 }_{ -0.01027}$  \\ 
 $\mMOM$  & 3 & $ 0.13085^{ +0.01162 }_{ -0.01174}$  \\ 
  & 4 & $ 0.13081^{ +0.01159 }_{ -0.01169}$  \\ 
 \hline 
  & 2 & $ 0.12650^{ +0.01048 }_{ -0.01031}$  \\ 
 $\MOMq$ & 3 & $ 0.13216^{ +0.01203 }_{ -0.01230}$  \\ 
  & 4 & $ 0.13024^{ +0.01137 }_{ -0.01137}$ \\ 
 \hline 
   & 2 & $ 0.12637^{ +0.01045 }_{ -0.01028}$  \\ 
 $\MOMg$ & 3 & $ 0.13115^{ +0.01171 }_{ -0.01187}$  \\ 
  & 4 & $ 0.13077^{ +0.01157 }_{ -0.01166}$  \\ 
 \hline 
  & 2 & $ 0.12677^{ +0.01054 }_{ -0.01039}$  \\ 
 $\MOMc$  &  3 & $ 0.13201^{ +0.01203 }_{ -0.01237}$  \\ 
  & 4 & $ 0.13059^{ +0.01150 }_{ -0.01156}$  \\ 
 \hline \hline 
 & 2 & $0.12680 \pm 0.00045^{ +0.01066 }_{ -0.01027}$\\ 
 Average & 3 & $0.13099 \pm 0.00117^{ +0.01203 }_{ -0.01124}$ \\ 
 & 4& $0.13053 \pm 0.00028^{ +0.01159 }_{ -0.01137}$\\ \hline 
 \end{tabular} 
  \end{center}\caption{{Estimates of $\alpha_s^{\MSbars} (M_Z)$ from the 
  R ratio effective coupling at order $\mbox{\L}$ using data from \cite{75} 
  calculated 
  in the $\MSbar$, $\MOMi$ and $\mMOM$ schemes. The error on the average is the 
  envelope of the scheme values and the average value is the centre of 
  this envelope.} 
\label{tab:RRDATA3}}
 \end{table}

\begin{figure}[ht]
\includegraphics[width=7.6cm,height=8cm]{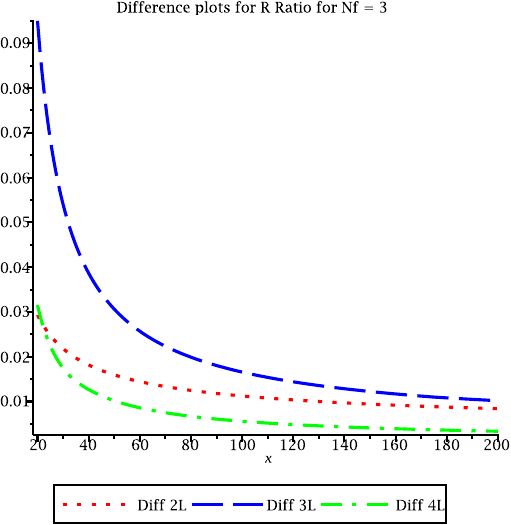}
\quad
\includegraphics[width=7.6cm,height=8cm]{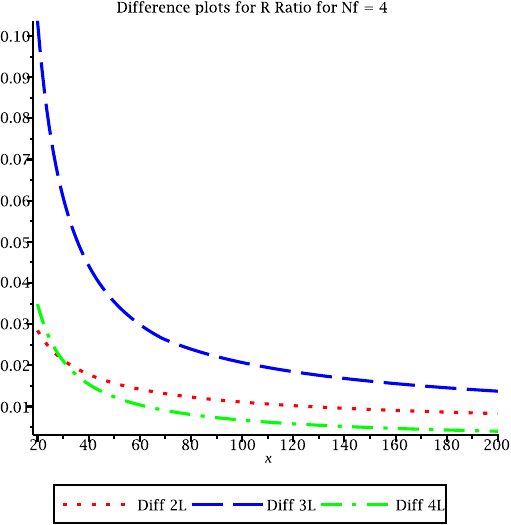}

\vspace{1.2cm}
\includegraphics[width=7.6cm,height=8cm]{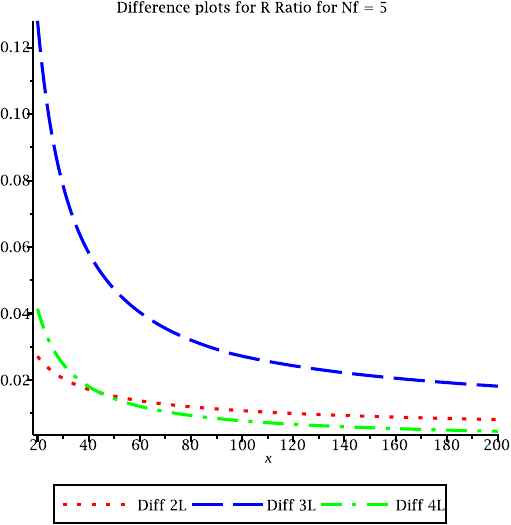}
\quad
\includegraphics[width=7.6cm,height=8cm]{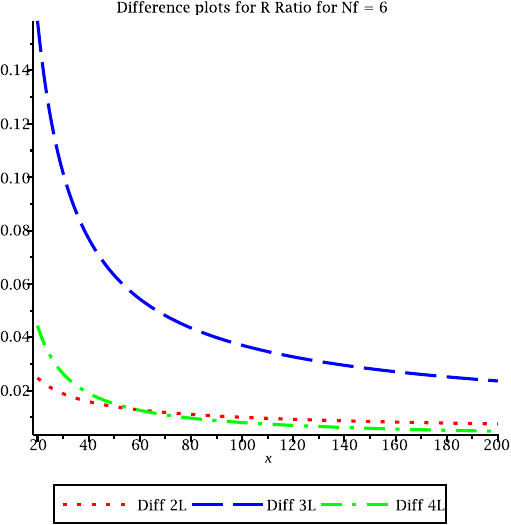}
\caption{Difference error for $a_{RR}^{\cal S}(x)$ at different 
loop orders for $\Nf$~$=$~$3$, $4$, $5$ and $6$ for the schemes 
considered here.}
\label{fig:DiffRRLoop}
\end{figure}

Instead of considering the envelopes themselves we can examine the difference 
between the maximum and minimum values in the envelope, as normalized by the 
midpoint of the envelope. This will provide a clearer display of the remaining 
scheme dependence. This is presented in Figure \ref{fig:DiffRRLoop} for the R 
ratio with Figure \ref{fig:DiffBJLoop} giving the corresponding situation for
the Bjorken sum rule. As discussed earlier, for the R ratio we see in general 
the scheme difference at $O(a^3)$ is largest, at $O(a^4)$ it is smallest for 
most of the range considered except at very low momenta where the $O(a^2)$
difference is smaller. For example, at $x$~$=$~$110$ for $\Nf$~$=$~$5$ the 
$O(a^2)$ difference is $1.0\%$, the $O(a^3)$ is $2.5\%$ and the $O(a^4)$ is 
$0.7\%$. The Bjorken sum rule provides a more conventional error for a 
perturbative series where in general the $O(a^2)$ error is larger than the 
$O(a^3)$ error which in turn is larger than the $O(a^4)$ error. 

 \begin{table}[ht]
\begin{center}
 \begin{tabular}{ |c|c|c| }
 \hline 
 \multicolumn{3}{|c|}{$\alpha_s^{{\mbox{\footnotesize{Bjr}}}}=0.70800 \pm 0.25716$}\\ 
 \multicolumn{3}{|c|}{at $Q= 1.64500$ GeV } \\ 
 \hline 
 \hline 
 &&\\ Scheme & $\mbox{\L}$  &$\alpha^{\MSbars}(M_Z)$ \\ & & \\ 
 \hline  \rule{0pt}{12pt} 
  & 2 & $ 0.12819^{ +0.01064 }_{-0.00538}$  \\ 
  $\MSbar$ & 3 & $ 0.12665^{ +0.01101 }_{-0.00574}$  \\ 
  & 4 & $ 0.12320^{ +0.00973 }_{-0.00498}$ \\ 
 \hline 
  & 2 & $ 0.12228^{ +0.00847 }_{-0.00403}$  \\ 
 $\mMOM$  & 3 & $ 0.12395^{ +0.00952 }_{-0.00536}$  \\ 
  & 4 & $ 0.12025^{ +0.00749 }_{-0.00328}$  \\ 
 \hline 
  & 2 & $ 0.12080^{ +0.00795 }_{-0.00374}$  \\ 
 $\MOMq$ & 3 & $-$ \\ 
  & 4 & $ 0.11752^{ +0.00659 }_{-0.00291}$ \\ 
 \hline 
   & 2 & $ 0.12340^{ +0.00823 }_{-0.00388}$  \\ 
 $\MOMg$ & 3 & $-$ \\ 
  & 4 & $ 0.12185^{ +0.00700 }_{-0.00279}$  \\ 
 \hline 
  & 2 & $ 0.12003^{ +0.00768 }_{-0.00364}$  \\ 
 $\MOMc$  &  3 & $-$ \\ 
  & 4 & $ 0.11752^{ +0.00614 }_{-0.00229}$  \\ 
 \hline \hline 
 & 2 & $0.12411 \pm 0.00408^{ +0.01064 }_{-0.00364}$ \\ Average & 3 & $\left[\frac{}{}0.12530 \pm 0.00135^{ +0.00536}_{-0.01101}\right]$ \\ & 4& $0.12036 \pm 0.00284^{ +0.00973 }_{-0.00229}$ \\ \hline \end{tabular} 
\end{center}\caption{{Estimates of $\alpha_s^{\MSbars} (M_Z)$ from the
 Bjorken sum rule effective coupling at order $\mbox{\L}$ using data from 
 \cite{76,77} calculated
 in the $\MSbar$, $\MOMi$ and $\mMOM$ schemes. The error on the average
 is the envelope of the scheme values and the average value is the centre
 of this envelope. The bracketed error indicates an envelope formed with the
 incomplete set of schemes.} \label{tab:BJDATA1}}
 \end{table}

\subsection{Estimating formal parameters.}

As a final quantitative scheme comparison we consider the fundamental formal 
parameters of the theory calculated using experimental data. In massless
perturbative QCD the $\Lambda$-parameter or equivalently the coupling constant 
defined at a particular renormalization scale are the only parameters that must
be inserted in order to make physical calculations. One can use the 
$\Lambda$-ratio or the coupling constant conversion functions to compare 
quantities calculated in each scheme in order to investigate residual scheme 
dependence. The most often quoted value is the coupling constant in the 
$\MSbar$ scheme evaluated at the mass of the $Z$ boson. Therefore that 
parameter, $\alpha_s^{\MSbars}(M_Z)$, will be considered here. We note that in
Nature resonances in partial interactions at any mass scale will be present and
reflected in associated experimental data. Such phenomena are not accounted for
in a massless theory. Therefore this study should also be regarded as part of a 
theory laboratory where we are particularly interested and can focus on the 
scheme dependence of the final result in a controlled setup rather than the 
precise value itself.

 \begin{table}[ht]
\begin{center}
 \begin{tabular}{ |c|c|c| }
 \hline 
 \multicolumn{3}{|c|}{$\alpha_s^{{\mbox{\footnotesize{Bjr}}}}=0.61700 \pm 0.25410$}\\ 
 \multicolumn{3}{|c|}{at $Q= 1.79500$ GeV } \\ 
 \hline 
 \hline 
 &&\\ Scheme & $\mbox{\L}$  &$\alpha^{\MSbars}(M_Z)$ \\ & & \\ 
 \hline  \rule{0pt}{12pt} 
  & 2 & $ 0.12736^{ +0.01470 }_{-0.00686}$  \\ 
  $\MSbar$ & 3 & $ 0.12559^{ +0.01488 }_{-0.00726}$  \\ 
  & 4 & $ 0.12242^{ +0.01335 }_{-0.00631}$ \\ 
 \hline 
  & 2 & $ 0.12194^{ +0.01211 }_{-0.00522}$  \\ 
 $\mMOM$  & 3 & $ 0.12331^{ +0.01315 }_{-0.00640}$  \\ 
  & 4 & $ 0.12019^{ +0.01106 }_{-0.00437}$  \\ 
 \hline 
  & 2 & $ 0.12057^{ +0.01147 }_{-0.00485}$  \\ 
 $\MOMq$ & 3 & $-$ \\ 
  & 4 & $ 0.11762^{ +0.00984 }_{-0.00384}$ \\ 
 \hline 
   & 2 & $ 0.12319^{ +0.01188 }_{-0.00503}$  \\ 
 $\MOMg$ & 3 & $ 0.12863^{ +0.01538 }_{-0.02879}$  \\ 
  & 4 & $ 0.12210^{ +0.01077 }_{-0.00384}$  \\ 
 \hline 
  & 2 & $ 0.11986^{ +0.01112 }_{-0.00470}$  \\ 
 $\MOMc$  &  3 & $-$ \\ 
  & 4 & $ 0.11784^{ +0.00959 }_{-0.00321}$  \\ 
 \hline \hline 
 & 2 & $0.12361 \pm 0.00375^{ +0.01470 }_{-0.00470}$ \\ Average & 3 & $\left[\frac{}{}0.12597 \pm 0.00266^{ +0.01538}_{-0.02347}\right]$ \\ & 4& $0.12002 \pm 0.00240^{ +0.01335 }_{-0.00384}$ \\ \hline \end{tabular} 
\end{center}\caption{{Estimates of $\alpha_s^{\MSbars} (M_Z)$ from the
 Bjorken sum rule effective coupling at order $\mbox{\L}$ using data from 
 \cite{76,77} calculated
 in the $\MSbar$, $\MOMi$ and $\mMOM$ schemes. The error on the average
 is the envelope of the scheme values and the average value is the centre
 of this envelope. The bracketed error indicates an envelope formed with the
 incomplete set of schemes.} \label{tab:BJDATA2}}
 \end{table}

We will use an experimental measurement for one of our perturbative series 
$\rho^{*}$ which was found at a particular energy level $Q^{*}$. With the 
series evaluated in a given scheme we can numerically solve for the 
$\Lambda$-parameter by varying $\Lambda^{\cal S}$ in 
\begin{equation}
\rho^{*} ~=~ \rho^{\cal S}({Q^{*}},{\Lambda^{\cal S}_{\Nf}}) ~.
\end{equation}
If in running from $Q^{*}$ to $M_Z$ we cross any threshold $T$ that exists in 
the massive theory we still consider the massless case but we change the number
of active fermions. For example, say there are $\Nf$ active fermions at $Q^{*}$
but in running to $M_Z$ we cross the mass threshold where the next quark 
becomes active at mass $M_{T}$ then we have to accommodate $\Nf+1$ active
quarks in the analysis beyond $M_T$. We can find the value of the coupling 
constant at the threshold explicitly by substituting $M_{T}$ and the value 
found for $\Lambda^{\cal S}_{\Nf}$ into the explicit formula for the coupling 
constant given in (\ref{eq:CC1L}). We consider the coupling constant to change
continuously at the threshold and therefore we solve  
\begin{eqnarray}
a^{\cal S}(M_{T},\Lambda^{\cal S}_{\Nf}) ~=~
a^{\cal S}(M_{T},\Lambda^{\cal S}_{\Nf+1}) 
\end{eqnarray}
numerically for $\Lambda^{\cal S}_{\Nf+1}$. This process is repeated until 
there are five current active fermions which is the number of active fermions 
at $M_Z$. We note that a similar process could be used to move down below a 
mass threshold as well.  Finally, we evaluate 
$a^{\cal S}(M_{Z}/\Lambda^{\cal S}_{5})$ explicitly and substitute the value 
into the coupling constant conversion function in order to have a value to
compare between the schemes.

 \begin{table}[ht]
\begin{center}
 \begin{tabular}{ |c|c|c| }
 \hline 
 \multicolumn{3}{|c|}{$\alpha_s^{{\mbox{\footnotesize{Bjr}}}}=0.58100 \pm 0.22308$}\\ 
 \multicolumn{3}{|c|}{at $Q= 1.96700$ GeV } \\ 
 \hline 
 \hline 
 &&\\ Scheme & $\mbox{\L}$  &$\alpha^{\MSbars}(M_Z)$ \\ & & \\ 
 \hline  \rule{0pt}{12pt} 
  & 2 & $ 0.12812^{ +0.01423 }_{-0.00711}$  \\ 
  $\MSbar$ & 3 & $ 0.12620^{ +0.01433 }_{-0.00747}$  \\ 
  & 4 & $ 0.12310^{ +0.01289 }_{-0.00650}$ \\ 
 \hline 
  & 2 & $ 0.12284^{ +0.01180 }_{-0.00546}$  \\ 
 $\mMOM$  & 3 & $ 0.12412^{ +0.01274 }_{-0.00649}$  \\ 
  & 4 & $ 0.12119^{ +0.01084 }_{-0.00464}$  \\ 
 \hline 
  & 2 & $ 0.12151^{ +0.01120 }_{-0.00508}$  \\ 
 $\MOMq$ & 3 & $-$ \\ 
  & 4 & $ 0.11866^{ +0.00966 }_{-0.00406}$ \\ 
 \hline 
   & 2 & $ 0.12417^{ +0.01161 }_{-0.00528}$  \\ 
 $\MOMg$ & 3 & $ 0.12907^{ +0.01453 }_{-0.03158}$  \\ 
  & 4 & $ 0.12330^{ +0.01067 }_{-0.00417}$  \\ 
 \hline 
  & 2 & $ 0.12081^{ +0.01087 }_{-0.00492}$  \\ 
 $\MOMc$  &  3 & $-$ \\ 
  & 4 & $ 0.11896^{ +0.00948 }_{-0.00350}$  \\ 
 \hline \hline 
 & 2 & $0.12446 \pm 0.00365^{ +0.01423 }_{-0.00492}$ \\ Average & 3 & $\left[\frac{}{}0.12660 \pm 0.00248^{ +0.01453 }_{-0.02663}\right]$ \\ & 4& $0.12098 \pm 0.00232^{ +0.01067 }_{-0.00406}$ \\ \hline \end{tabular} 
\end{center}\caption{{Estimates of $\alpha_s^{\MSbars} (M_Z)$ from the
 Bjorken sum rule effective coupling at order $\mbox{\L}$ using data from 
 \cite{76,77} calculated
 in the $\MSbar$, $\MOMi$ and $\mMOM$ schemes. The error on the average
 is the envelope of the scheme values and the average value is the centre
 of this envelope. The bracketed error indicates an envelope formed with the
 incomplete set of schemes.} \label{tab:BJDATA3}}
 \end{table}
 
In practice we have utilised the $\Lambda$-ratio and a numerical root finding 
algorithm to search for the value of $\Lambda^{\MSbars}$ which would result in 
the correct value of the R ratio at a given energy level. A fixed range for 
$\Lambda^{\MSbars}$ of $50-800$ MeV has been chosen to cover much of the 
perturbative regime, where we expect $\Lambda^{\MSbars}$~$\sim$~$200$ MeV
\cite{70}. A root bisection algorithm was used to find the value of $y$ that 
gives the solution of 
\begin{eqnarray}
\rho^{*}-\rho^{\cal S}\Bigg(Q^{*}\frac{\Lambda^{\MSbars}}
{\Lambda^{\cal S}},y\Bigg) ~=~ 0 ~.
\end{eqnarray}
where ${\cal S}$ may be any of the schemes considered including $\MSbar$ in
which the above $\Lambda$-ratio to the $\MSbar$ scheme is unity. As the 
numerical $\Lambda$ value is intermediate to our analysis care was taken to 
ensure the result was sufficiently accurate for further calculations.

 \begin{table}[ht]
\begin{center}
 \begin{tabular}{ |c|c|c| }
 \hline 
 \multicolumn{3}{|c|}{$\alpha_s^{{\mbox{\footnotesize{Bjr}}}}=0.63600 \pm 0.18702$}\\ 
 \multicolumn{3}{|c|}{at $Q= 2.17700$ GeV } \\ 
 \hline 
 \hline 
 &&\\ Scheme & $\mbox{\L}$  &$\alpha^{\MSbars}(M_Z)$ \\ & & \\ 
 \hline  \rule{0pt}{12pt} 
  & 2 & $ 0.13287^{ +0.00957 }_{-0.00564}$  \\ 
  $\MSbar$ & 3 & $ 0.13093^{ +0.00979 }_{-0.00594}$  \\ 
  & 4 & $ 0.12741^{ +0.00866 }_{-0.00514}$ \\ 
 \hline 
  & 2 & $ 0.12691^{ +0.00768 }_{-0.00429}$  \\ 
 $\mMOM$  & 3 & $ 0.12852^{ +0.00853 }_{-0.00521}$  \\ 
  & 4 & $ 0.12498^{ +0.00685 }_{-0.00360}$  \\ 
 \hline 
  & 2 & $ 0.12542^{ +0.00722 }_{-0.00398}$  \\ 
 $\MOMq$ & 3 & $-$ \\ 
  & 4 & $ 0.12215^{ +0.00601 }_{-0.00314}$ \\ 
 \hline 
   & 2 & $ 0.12826^{ +0.00750 }_{-0.00414}$  \\ 
 $\MOMg$ & 3 & $ 0.13502^{ +0.01105 }_{-0.02911}$  \\ 
  & 4 & $ 0.12715^{ +0.00657 }_{-0.00320}$  \\ 
 \hline 
  & 2 & $ 0.12464^{ +0.00698 }_{-0.00386}$  \\ 
 $\MOMc$  &  3 & $-$ \\ 
  & 4 & $ 0.12235^{ +0.00571 }_{-0.00263}$  \\ 
 \hline \hline 
 & 2 & $0.12876 \pm 0.00411^{ +0.00957 }_{-0.00386}$ \\ Average & 3 & $\left[\frac{}{}0.13177 \pm 0.00325^{ +0.01105 }_{-0.00325}\right]$ \\ & 4& $0.12478 \pm 0.00263^{ +0.00866 }_{-0.00314}$ \\ \hline \end{tabular} 
\end{center}\caption{{Estimates of $\alpha_s^{\MSbars} (M_Z)$ from the
 Bjorken sum rule effective coupling at order $\mbox{\L}$ using data from 
 \cite{76,77} calculated
 in the $\MSbar$, $\MOMi$ and $\mMOM$ schemes. The error on the average
 is the envelope of the scheme values and the average value is the centre
 of this envelope. The bracketed error indicates an envelope formed with the
 incomplete set of schemes.} \label{tab:BJDATA4}}
 \end{table}

We recall that the world average value of the effective coupling constant at 
the mass of the $Z$ boson is 
$\alpha_s^{\MSbars}(M_Z)$~$=$~$0.1179$~$\pm$~$0.0009$ \cite{78} where 
$\alpha_s^{\MSbars}$~$=$~$4\pi a_{\MSbars}$. Since this coupling constant is 
more commonly used we have chosen to display our results with this convention. 
Converting the R ratio data given in \cite{75} to our effective coupling 
constant and using the above described method for finding 
$\alpha_s^{\MSbars}(M_Z)$ in each scheme with the data our results are recorded
in Tables \ref{tab:RRDATA1}, \ref{tab:RRDATA2} and \ref{tab:RRDATA3} which use
experimental data from \cite{75}. The values given at each loop order in each 
scheme are for the central experimental value. The errors are found by solving 
for the upper and lower bounds on the experimental value. We have added an 
additional row for each loop order giving the average value of the other scheme
values at that loop order. As a scheme distribution is not known we have chosen
to use the central value of the envelopes as the average with the first error 
being the theory error from the scheme difference envelope and the upper and 
lower bound on the second error being the experimental error on the maximum and
minimum central value from the schemes, respectively.  

In Tables \ref{tab:RRDATA1}, \ref{tab:RRDATA2} and \ref{tab:RRDATA3} we see the
general trend discussed before where the scheme error at two loops increases 
drastically at three loops and then decreases below the two loop error at four 
loops. We note that the values calculated for $\alpha_s^{\MSbars}$ are larger 
than the current world average. However since we are considering massless QCD 
some difference is expected and the values found here are commensurate with 
similar analysis made using the same data in \cite{49}. Despite our unphysical 
assumptions, this analysis does show that the general trend of reduced scheme 
dependence at the four loop level can translate to reduced theory error in 
formal quantities calculated with experimental data. On general field 
theoretical grounds this is not unexpected but the analysis has provided a 
degree of quantification.

\begin{figure}[ht]
\includegraphics[width=7.6cm,height=8cm]{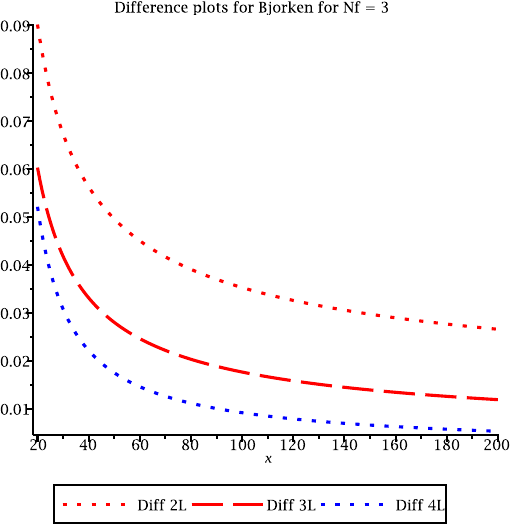}
\quad
\includegraphics[width=7.6cm,height=8cm]{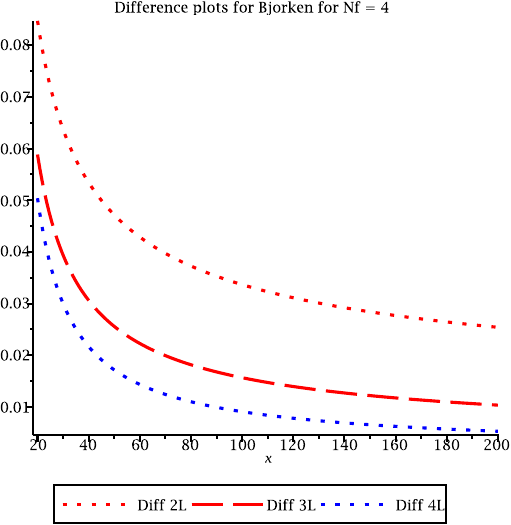}

\vspace{1.2cm}
\includegraphics[width=7.6cm,height=8cm]{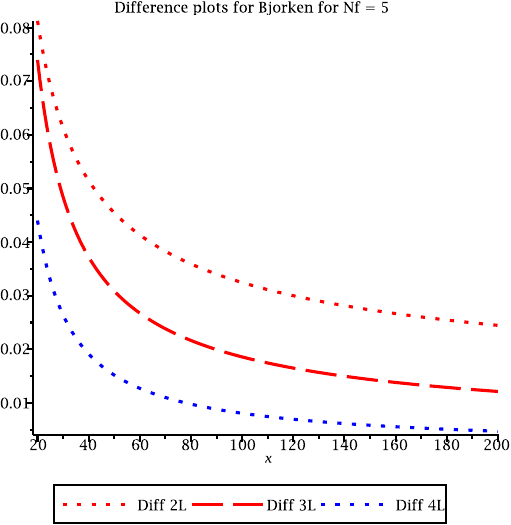}
\quad
\includegraphics[width=7.6cm,height=8cm]{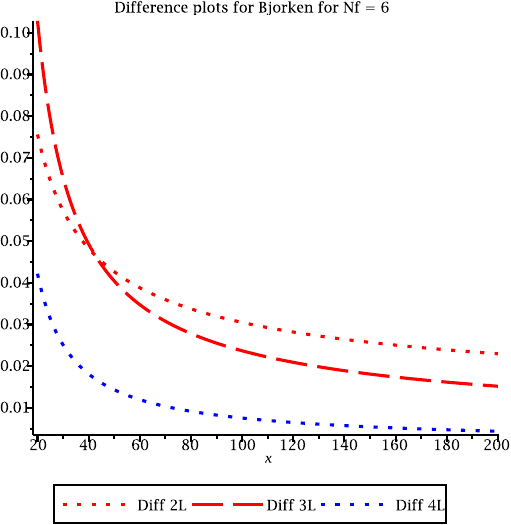}
\caption{Difference error for $a_{RR}^{\cal S}(x)$ at different 
loop orders for $\Nf$~$=$~$3$, $4$, $5$ and $6$ for the schemes 
considered here.}
\label{fig:DiffBJLoop}
\end{figure}

We have repeated this exercise for the Bjorken sum rule using data from
\cite{76,77} with the results given in Tables \ref{tab:BJDATA1},
\ref{tab:BJDATA2}, \ref{tab:BJDATA3} and \ref{tab:BJDATA4}. In this case we 
used the higher $Q^2$ values of the effective coupling constant presented in 
Table I of \cite{46} derived from the deeply virtual Compton scattering 
measurement of the sum rule recorded in \cite{76,77}. Due to the low energies 
considered in this case these data points push the boundaries of validity for 
the perturbative regime. Therefore a three loop estimate could not be made for 
one or more instances for the symmetric momentum subtraction schemes as their 
three loop series do not meet the experimental effective coupling constant 
value. Although, this reduced scheme error at three loops compared with the two
loop result is consistent with our difference plots of the running of the 
Bjorken sum rule in different schemes provided in Figure \ref{fig:DiffBJLoop}.
The smaller error on the three loop average values can then be understood as 
due to the reduction in the schemes considered, especially since the $\MSbar$ 
and $\mMOM$ ones are related as described earlier. Importantly though the 
scheme dependence is again reduced at the four loop level. Indeed for the 
higher $Q^2$ values in the Bjorken case the four loop error for what is termed 
the average, which includes the MOM schemes, is compatible if not better than 
the three loop average. In other words the fact that all schemes provide 
results at four loops, despite not giving results at three loops, could be 
interpreted as the series settling down at the new loop level. However it may 
be the case that this is simply due to the fact this is at the interface of 
perturbative reliability for low loop order. So there is a large contribution 
at the four loop level which is positive in all cases meaning the term that 
will dominate as the $\Lambda$-parameter is decreased will increase the value 
of the effective coupling constant which in turn means the series will increase
above the experimental values. Overall the values found from the Bjorken data 
are much lower than those of the R ratio, with the Bjorken data being closer to
the current world average. 
 
\sect{Discussion.}

We have completed a comprehensive investigation into the scheme dependence of
two observables at as high a loop order as is presently possible. This included
the kinematic schemes of \cite{20,21} which are based on a specific momentum
configuration of the core $3$-point vertices of QCD. Our main aim was to 
quantify the error on a benchmark parameter, $\alpha_s^{\MSbars}(M_Z)$, that
was more field theoretically based and accounted for the uncertainty that is
inherent in the truncation of the perturbation series in different schemes. We
stress that our formulation was in the idealized and purely theoretical
situation where quark masses and threshold effects were not included. We remark
parenthetically, however, that in the former instance a hybrid or partial MOM 
scheme investigation could be instigated in the approximation of heavy mass 
corrections, such as \cite{79} which includes the Bjorken sum rule, since the 
coupling constant and quark mass, treated as parameters, can be mapped from the
$\MSbar$ scheme to one of the $\MOMi$ ones from available data. The 
construction is termed hybrid or partial in the sense that the $\MOMi$ scheme 
maps do not include quark masses but would have an approximation in a heavy 
quark mass. Further, while recognizing that our massless investigation is not 
realistic on the contrary it allowed us to focus purely on the effect 
corrections in different schemes have in a controlled scenario. Although the 
central value for $\alpha_s^{\MSbars}(M_Z)$ differed for the two cases of the 
Bjorken sum rule and the R ratio what was generally apparent is that with 
increasing loop order there was a narrowing of our error estimates and an 
indication of convergence. For the various schemes we concentrated on this 
improvement was prevalent at four and higher orders which is encouraging. 
Moreover that in itself justifies the need to progress the renormalization of 
QCD in $\MOMi$ schemes to the next order. To improve our error analysis further
more data and additional schemes could be considered. Here we concentrated on 
the $\mMOM$, $\MSbar$ and $\MOMi$ schemes, many of which are related via the 
renormalization group construction meaning that the final results from each 
scheme cannot be considered as independent. Indeed they are very much connected
via the underlying properties of the renormalization group equation. In 
addition to this much of the data considered is outside the range where 
perturbation theory is ordinarily applied, particularly for the Bjorken sum 
rule. So considering more data in the perturbative regime should provide more 
accurate information on scheme dependence. Additional investigation of the 
scheme distribution could further inform our results allowing for a better 
average and degree of belief to be attached to our error estimates.

Aside from the fully symmetric point schemes of \cite{20,21} considered here 
there are extensions of those to more general kinematic schemes. For instance, 
one variation of \cite{20,21} was introduced in \cite{80,81} where a parameter 
$\omega$ reflected the relative weighting of the external momentum flowing 
through one of the legs of the $3$-point vertices. Such a parameter would 
naturally translate to a suite of schemes generalizing those of \cite{20,21}. 
As the range of $\omega$ is limited to $\omega$~$\in$~$(0,4)$, with the 
bounding values originating from infrared or colinear singularities, then 
$\omega$ could provide a more natural way to tune or quantify the range of 
theoretical errors deriving from an application of the approach discussed here.
Equally rather than have one controlling parameter we recall that there are 
three independent variables for an off-shell $3$-point vertex. These are one 
overall scale and two dimensionless momenta ratios which, like $\omega$, are 
bounded but in this instance the dimensionless variables are constrained to lie
within a paraboloid. While this too could translate to a bounding region on 
theory errors of an observable only the underlying two loop massless off-shell 
master integrals are available at present \cite{28,62,64,65,82}. Knowledge of 
the three loop fully off-shell masters would be needed before a concrete 
analysis of this extensions could proceed. Moreover if achieved it would be of 
interest to compare with other methods of extracting scale independent 
information from observables. Ultimately it ought to be the case that with high
enough loop order information all well-founded methods of determining scale 
independent results for an observable should themselves converge to the same 
value. Another direction that was briefly considered in \cite{29} was the
inclusion of the gauge parameter in the evolution of the effective coupling
derived from the R ratio in the MOM schemes. Such dependence could equally be
embedded in an error analysis. However before that could proceed to the order
considered here the renormalization group functions in the three MOM schemes
would have to be determined for non-zero $\alpha$. Currently only the four loop
Landau gauge expressions are available \cite{24}. Finally, in the more 
immediate future extending the present work to the five loop level would 
provide further insight into whether the scale of reduction in scheme 
dependence is due to an artifact of the underlying scheme used at the four loop
level or because it is due to true scheme independence. In the short term it 
would seem that all the technology to analytically compute the four loop 
massless symmetric $3$-point master Feynman integrals is actually available. 
For instance, the first step of carrying out the momentum expansion of the four
loop masters to high order is viable now that the four loop {\sc Forcer} 
algorithm, \cite{83,84}, written in {\sc Form}, \cite{60,61}, has superseded 
the three loop {\sc Mincer} package. 

\vspace{1cm}
\noindent
{\bf Acknowledgements.} This work was supported in part by a DFG Mercator 
Fellowship (JAG), STFC Consolidated Grant ST/T000988/1 (JAG) and an EPSRC 
Studentship EP/R513271/1 (RHM). For the purpose of open access, the authors 
have applied a Creative Commons Attribution (CC-BY) licence to any Author 
Accepted Manuscript version arising. The electronic versions representing the 
Landau gauge expressions for the Bjorken sum rule and R ratio in the $\mMOM$ 
and $\MOMi$ schemes used here are accessible from the arXiv ancillary directory 
associated with the article. The symbolic manipulation language {\sc Form},
\cite{60,61}, was employed for various calculations in this article. RHM thanks
A. Freitas and T. Teubner for useful conversations.

{\begin{figure}[ht]
\includegraphics[width=7.6cm,height=8.00cm]{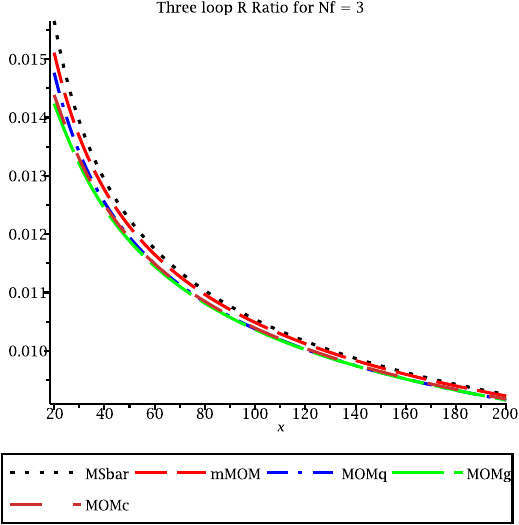}
\quad
\includegraphics[width=7.6cm,height=8.00cm]{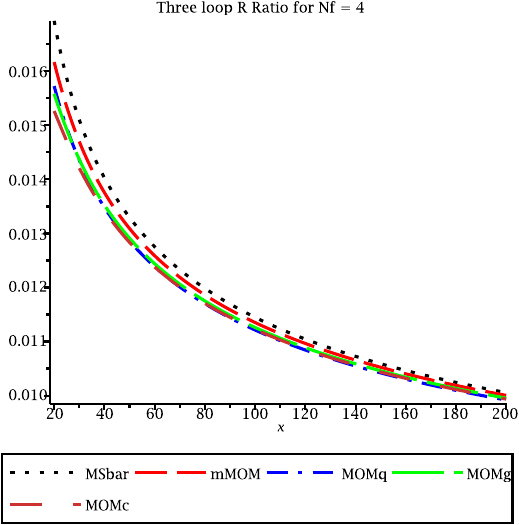}
\quad

\vspace{1.2cm}

\includegraphics[width=7.6cm,height=8.00cm]{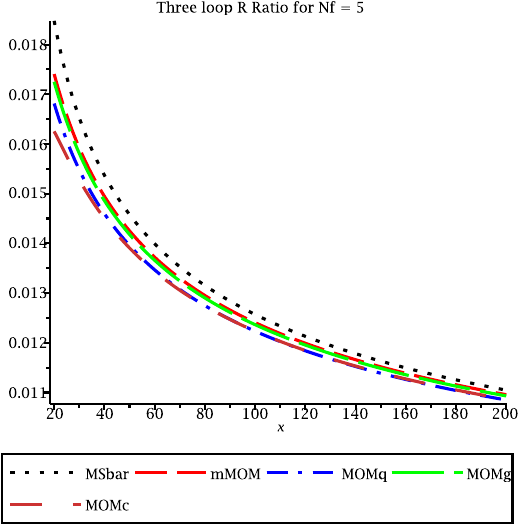}
\quad
\includegraphics[width=7.6cm,height=8.00cm]{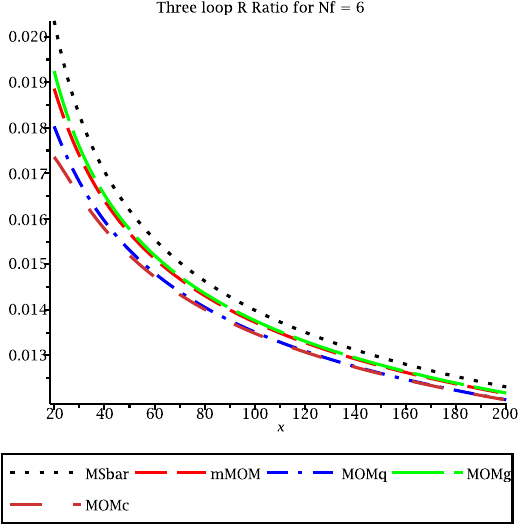}
\quad
\caption{Comparison of $a_{RR}^{\cal S}(x)$ at three loops for the various 
schemes for $\Nf$~$=$~$3$, $4$, $5$ and $6$.}\label{fig:RR3LScheme}
\end{figure}}

{\begin{figure}[ht]
\includegraphics[width=7.6cm,height=8cm]{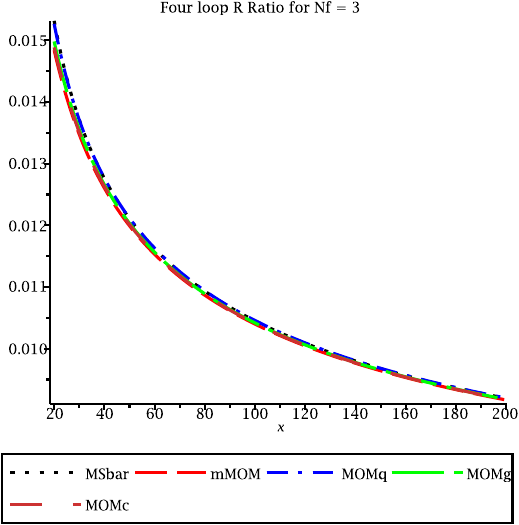}
\quad
\includegraphics[width=7.6cm,height=8cm]{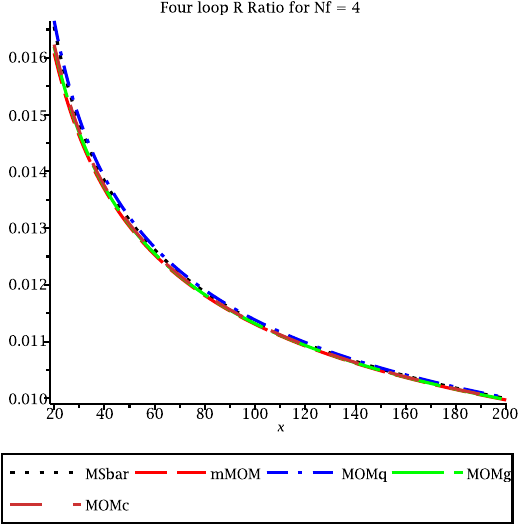}

\vspace{1.2cm}
\includegraphics[width=7.6cm,height=8cm]{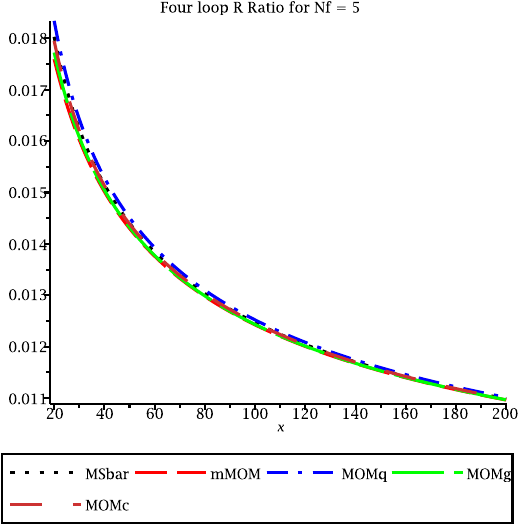}
\quad
\includegraphics[width=7.6cm,height=8cm]{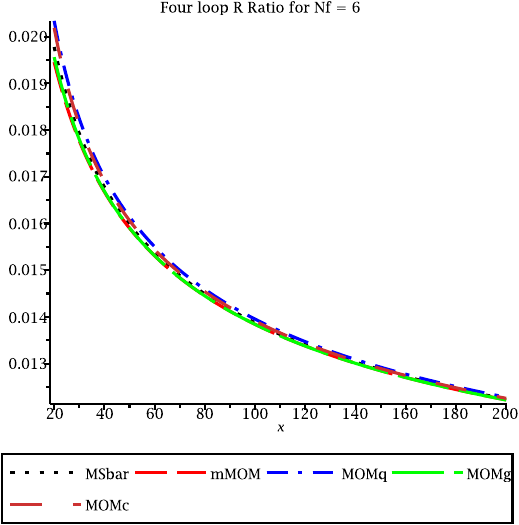}
\caption{Comparison of $a_{RR}^{\cal S}(x)$ at four loops for the various 
schemes for $\Nf$~$=$~$3$, $4$, $5$ and $6$.}\label{fig:RR4LScheme}
\end{figure}}

{\begin{figure}[ht]
\includegraphics[width=7.6cm,height=8cm]{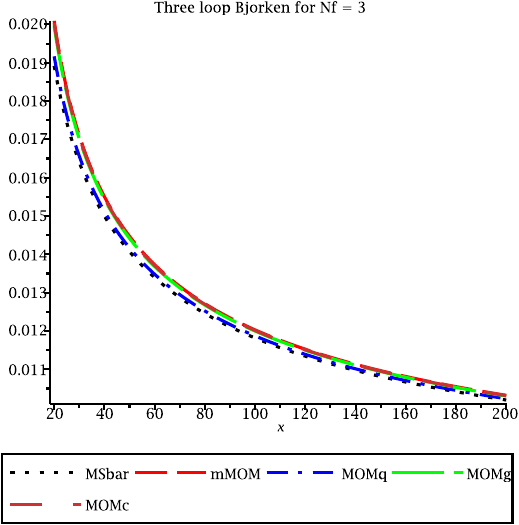}
\quad
\includegraphics[width=7.6cm,height=8cm]{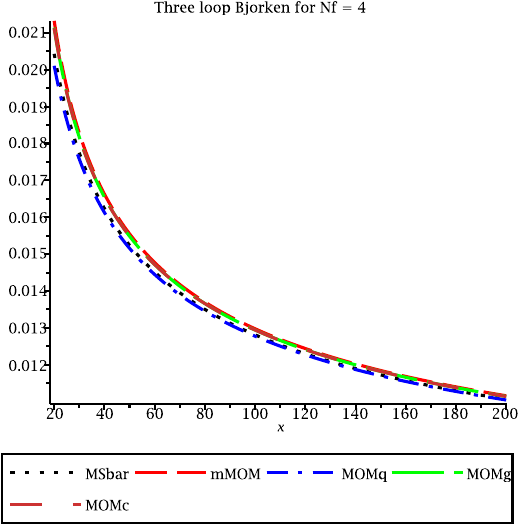}
\quad

\vspace{1.2cm}

\includegraphics[width=7.6cm,height=8cm]{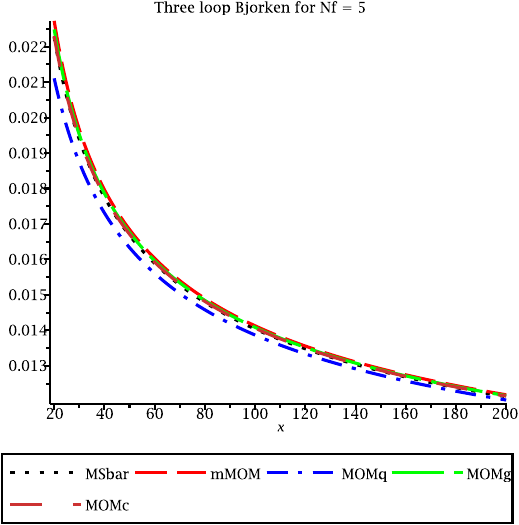}
\quad
\includegraphics[width=7.6cm,height=8cm]{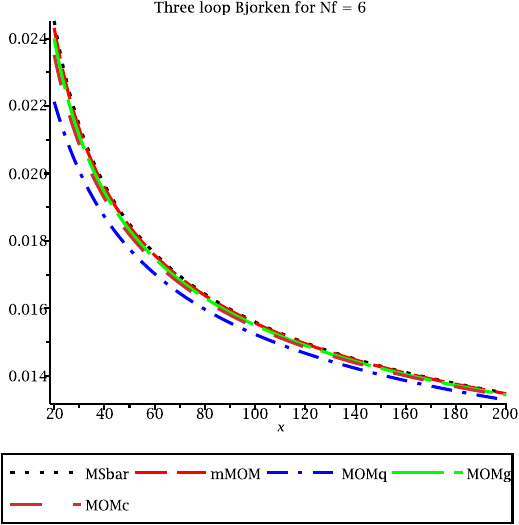}
\quad
\caption{Comparison of $a_{{\mbox{\footnotesize{Bjr}}}}^{\cal S}(x)$ at three loops for the various 
schemes for $\Nf$~$=$~$3$, $4$, $5$ and $6$.}\label{fig:BJ3LScheme}
\end{figure}}

{\begin{figure}[ht]
\includegraphics[width=7.6cm,height=8cm]{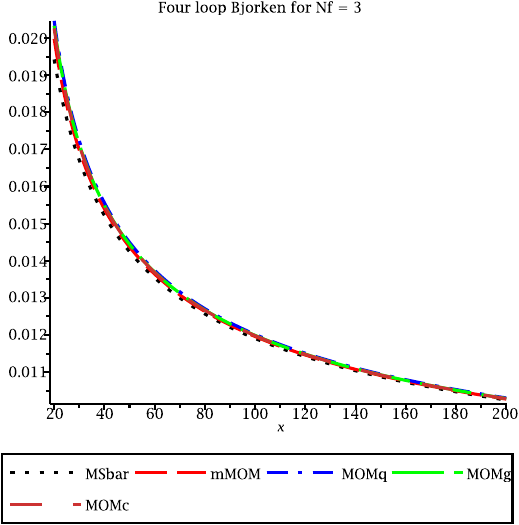}
\quad
\includegraphics[width=7.6cm,height=8cm]{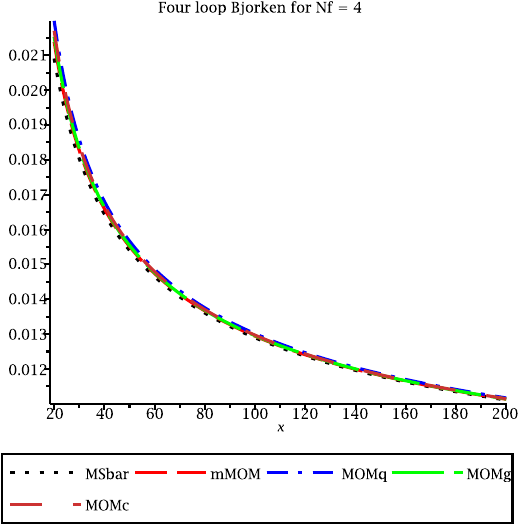}

\vspace{1.2cm}
\includegraphics[width=7.6cm,height=8cm]{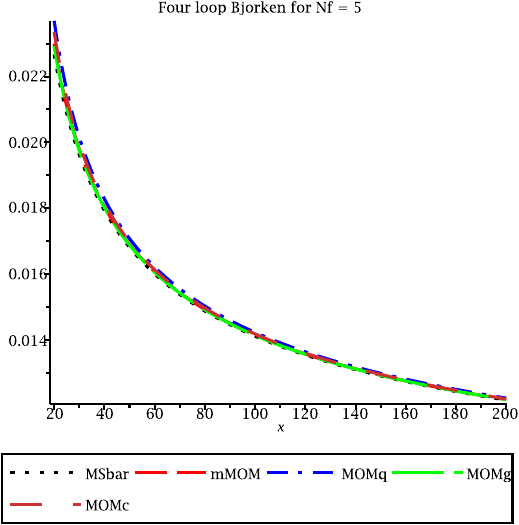}
\quad
\includegraphics[width=7.6cm,height=8cm]{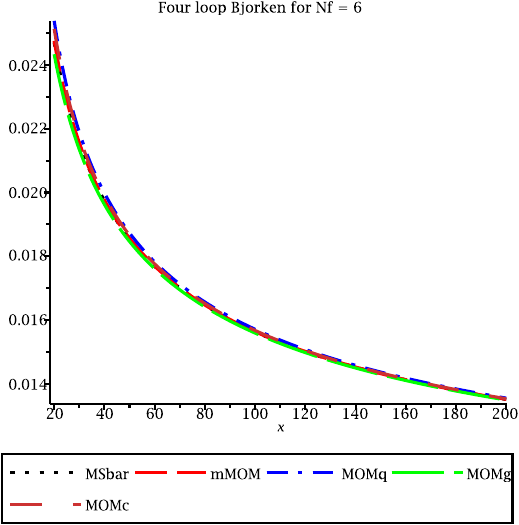}
\caption{Comparison of $a_{{\mbox{\footnotesize{Bjr}}}}^{\cal S}(x)$ at various
loops for the various schemes for $\Nf$~$=$~$3$, $4$, $5$ and $6$.}
\label{fig:BJ4LScheme}
\end{figure}}

\end{document}